 \documentclass[final,3p,times,twocolumn]{elsarticle}

\usepackage{lineno}
\usepackage[hidelinks]{hyperref}
\modulolinenumbers[5]

\journal{Renewable Energy}
\usepackage{commath,physics}
\usepackage[nameinlink,noabbrev]{cleveref}%
\usepackage{amsmath,todonotes}
\usepackage{graphicx}

\usepackage[locale=UK,
number-mode=math,
unit-mode=text,
per-mode=symbol,
number-unit-product=\ ,
retain-zero-exponent=false]{siunitx}[=v2]
\DeclareSIUnit{\pixel}{px}
\DeclareSIUnit{\fps}{fps}

\newcommand{\kindex}[2]{\ensuremath{{#1}_{\scalebox{0.65}{#2}}}}
\newcommand{\U}{\textrm{U}}
\newcommand{\Uinf}{\kindex{\U}{$\infty$}}

\newcommand{\Rey}{\mbox{Re}}
\newcommand{\subf}[1]{(#1)}

\graphicspath{{../../figurematter/latex/figs/}{../../figurematter/plots/}}

\begin{document}

\begin{frontmatter}

\title{The dynamic stall dilemma for vertical-axis wind turbines}

\author{S\'ebastien Le Fouest}
%\author{David Bensason} %\corref{newaffil}
\author{Karen Mulleners\corref{mycorrespondingauthor}}
\cortext[mycorrespondingauthor]{Corresponding author}
\ead{karen.mulleners@epfl.ch}

\address{Institute of Mechanical Engineering, \'Ecole Polytechnique F\'ed\'erale de Lausanne (EPFL), CH-1015 Lausanne, Switzerland}

\begin{abstract}
Vertical-axis wind turbines (VAWT) are excellent candidates to complement traditional wind turbines and increase the total wind energy capacity. Development of VAWT has been hampered by their low efficiency and structural unreliability, which are related to the occurrence of dynamic stall. Dynamic stall consists of the formation, growth, and shedding of large-scale dynamic stall vortices, followed by massive flow separation. The vortex shedding is detrimental to the turbine’s efficiency and causes significant load fluctuations that jeopardise the turbine’s structural integrity. We present a comprehensive experimental characterisation of dynamic stall on a VAWT blade including time-resolved load and velocity field measurements. Particular attention is dedicated to the dilemma faced by VAWT to either operate at lower tip-speed ratios to maximise their peak aerodynamic performance but experience dynamic stall, or to avoid dynamic stall at the cost of reducing their peak performance. Based on the results, we map turbine operating conditions to one of three regimes: deep stall, light stall, and no stall. The light stall regime offers VAWT the best compromise in the dynamic stall dilemma as it yields positive tangential forces during the upwind and downwind rotation and reduces load transients by \SI{75}{\percent} compared to the deep stall regime.
\end{abstract}


\end{frontmatter}

\newpage

%				--------				~			introduction		~					--------				%

\section{Introduction}

%				--------				~			introduction		~					--------				%

Wind energy is recognised as a potential primary source in a near-zero-emission global power production scheme \cite{Marvel2013,Petersen2017,Dupont2018}.
The potential of wind-generated electricity was assessed to be about seven times greater than the global electricity demand and about the same magnitude as the energy demand in all forms in 2017 \cite{Lu2009,Lu2017}.
Concerns around the logistical feasibility of the predicted exponential deployment rate of current wind turbine technology have challenged whether wind energy can truly meet this potential \cite{Hansen2017,Dupont2018}.
Physical and geographic constraints hamper wind energy availability and accessibility which could limit the energy return on investment and make other sources of energy more favourable than wind energy \cite{Dupont2018,Johansson2012,DeCastro2011,Miller2011}.
The diversification of wind energy technology has the potential to mitigate concerns around accessibility and give a boost to the contribution of wind energy to the global energy mix.

Vertical-axis wind turbines are excellent candidates for wind technology diversification \cite{Xie2017,Hezaveh2018,Dabiri2011}.
They feature several advantages that complement the energy conversion by their horizontal counterparts.
One of these features is the insensitivity to the wind direction, which allows H-type Darrieus wind turbines to prosper in highly sheared or even vortex dominated flows.
Compared to conventional horizontal-axis wind turbines, vertical-axis wind turbines have fewer moving parts, lower installation and maintenance costs as their generator is typically installed at ground level, and improved scalability, robustness, and lower noise level due to lower operational tip-speed ratios \cite{Rezaeiha2017,Buchner2018,Rolin2018}.
But, the development of efficient and reliable vertical-axis wind turbine design is limited by their aerodynamic complexity.

\begin{figure*}[tb]
\centering
\includegraphics[width=\linewidth]{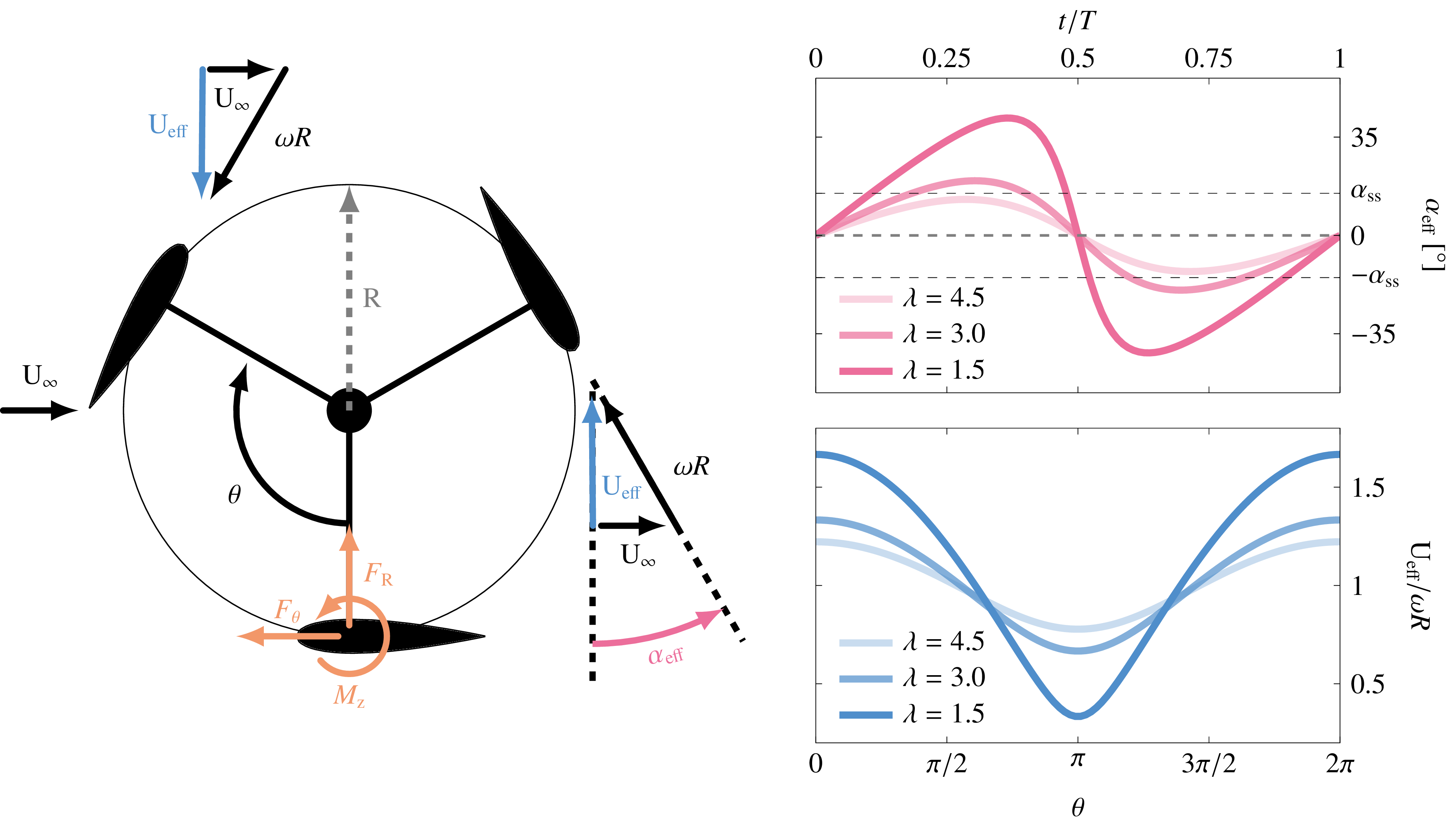}
\caption{Vertical-axis wind turbine blade kinematics.
The free stream velocity $\protect\Uinf$ goes from left to right.
Indication of the positive direction of the radial force \kindex{F}{R}, azimuthal force \kindex{F}{$\theta$}, and pitching moment at quarter-chord \kindex{M}{z}.
The blade's velocity is equivalent to the rotational frequency $\omega$ times the turbine's radius $R$.
Schematic representation of the definition of the blade's effective angle of attack $\kindex{\alpha}{eff}$ (\cref{eq:Aeff}) and velocity $\kindex{U}{eff}$ (\cref{eq:Ueff}) and their temporal evolution as a function of the blade's azimuthal position.
}
\label{fig:vawtaero}
\end{figure*}

The blades of a vertical-axis wind turbine rotate about an axis perpendicular to the free stream, such that the effective inflow velocity and angle of attack of the blade vary periodically over a rotation cycle (\cref{fig:vawtaero}).
The amplitude and skewness of this variation is governed by the tip-speed ratio $\lambda$, a dimensionless parameter obtained from the ratio between the blade velocity $\omega R$ and the free stream velocity \Uinf:
\begin{equation}
	\lambda = \frac{\omega R}{\Uinf}\quad.
	\label{eq:tsr}
\end{equation}
Here, $R$ is the rotor radius and $\omega$ is the rotational velocity of the rotor.
Assuming that the blade starts its rotation facing the wind, the variation of the angle of attack $\kindex{\alpha}{eff}$ with respect to the blade's azimuthal position $\theta$ and the tip-speed ratio $\lambda$ can be derived from trigonometry:
\begin{equation}
	\kindex{\alpha}{eff}(\theta) = \tan^{-1}\left(\frac{\sin \theta}{\lambda+\cos \theta}\right)\quad.
	\label{eq:Aeff}
\end{equation}
The effective flow velocity \kindex{\U}{eff} acting on the turbine blade can also be expressed relative to the incoming wind speed with respect to the blade's azimuthal position:
\begin{equation}
\kindex{\U}{eff}(\theta) = \Uinf \sqrt{1+2 \lambda \cos \theta+\lambda^{2}}\quad.
\label{eq:Ueff}
\end{equation}
Note that \cref{eq:Aeff,eq:Ueff} do not consider any wake-induced effects and are derived purely from the turbine' geometry and kinematics.

For vertical-axis wind turbines operating at low tip-speed ratios, typically $\lambda<3$, the blade undergoes large variations in effective flow conditions (\cref{fig:vawtaero}).
The large unsteady excursions of the effective angle of attack beyond its static stall angle $\kindex{\alpha}{ss}$ and the varying inflow velocity can lead to the occurrence of dynamic stall \cite{Fujisawa2001,Buchner2018,SimaoFerreira2009}.
Dynamic stall is characterised by the formation, growth, and shedding of large-scale vortices \cite{Carr1977}.
Massive flow separation is delayed to higher angles of attack which leads to a lift overshoot with respect to the static response.
Although these attributes appear beneficial, dynamic stall is generally not considered desirable.
The shedding of large-scale vortices is generally followed by a dramatic loss in aerodynamic efficiency and highly unsteady loads \cite{Mccroskey1982,Mulleners2013}.
When the blade's effective angle of attack exceeds the blade's critical stall angle in the upwind half of the rotation ($0 \leq \theta < \pi$) for low tip-speed ratios, stall vortices are formed on the inboard side of the blade.
When these vortices shed, they convect downstream and interact with the blade for an extended period of time throughout the downwind half of the blade's rotation ($\pi \leq \theta < 2\pi$) \cite{SimaoFerreira2009,Ferreira2009,Dave2021}.
Blade-vortex interaction and post-stall vortex shedding cause highly transient and heavy load fluctuations that jeopardise the turbine's structural integrity and often lead to premature fatigue failure \cite{DeTavernier2021,Dunne2015,Ouro2018}.
%Optimal operating tip-speed ratios for power production were found to be as low as $\lambda = 1$ by Miller et al.\cite{Miller2018}, who studied a wind turbine operating at full dynamic similarity.
The low tip-speed ratio operation is associated with high variations in angle of attack, dynamic stall, and extended periods of time during which the flow is massively separated, compared to higher tip-speed ratio operation.

%%%% todo start
The temporal development of dynamic stall and its influence on the aerodynamic loads on vertical-axis wind turbines has been investigated experimentally and numerically in the past decades.
Early experimental studies revealed that dynamic stall occurs for tip-speed speed ratio $\lambda<4$ for a chord-to-diameter ratio $c/D = 0.1$ and transitions from light stall to deep stall as the tip-speed ratio decreases \cite{Laneville1986}.
Among the first flow measurement of dynamic stall on vertical-axis wind turbines were performed by \cite{SimaoFerreira2009}.
They quantify the strength of the dynamic stall vortex which increases with decreasing tip-speed ratio.
The higher vortex strength is attributed to the higher angle of attack variations that the blade experiences at lower tip-speed ratios.
For higher tip-speed ratios, dynamic stall is delayed to later azimuthal positions and is less severe, with smaller and weaker vortical structures shed from the blade.
Spatially resolved stereoscopic particle-image velocimetry for tip-speed ratio $\lambda \in [\num{1}-\num{5}]$, Reynolds number $Re \in [\num{50000} - \num{140000}]$ and chord-to-diameter ratio $c/D \in [\num{0.1} - \num{0.2}]$ were presented in \cite{Buchner2018}.
The topology, timing, and strength of the dynamic stall vortex vary as a function of the tip-speed ratio and the chord-to-diameter ratio.
The tip-speed ratio dictates the amplitude of effective flow variations and the chord-to-diameter ratio governs the frequency of these variations.
The evolution of the vortex strengths for different chord-to-diameter ratios collapses when presented as a function of the convective time post-stall, i.e, the number of chords travelled by the blade after exceeding static stall.
This suggests that vortex formation is a self-similar across reduced frequencies and that vortex strength is governed only by the tip-speed ratio.

The goal of this paper is to provide a detailed analysis of the interplay between dynamic stall and the aerodynamic performance of a single-bladed vertical-axis wind turbine or cross-flow turbine blade.
This was achieved by experimentally collecting time-resolved unsteady aerodynamic loads and time-resolved flow field data on a scaled-down H-type Darrieus wind turbine operating at tip-speed ratios ranging from \numrange{1.2}{6}.
Successive stages in the flow development within a single turbine rotation are identified and their influence of the loads experienced by the blade are described for different tip-speed ratios.
Specific metrics are proposed to evaluate the performance of the wind turbine blade in terms of torque production and by the level of load transients related to flow separation.
The timescales related to dynamic stall development are systematically quantified and compared to universal vortex formation timescales that are previously identified in literature.
Based on the temporal evolution of unsteady loads, flow structures, and load transients, we establish criteria to categorise the turbine's operation into one of three regimes: no stall, light stall, and deep dynamic stall.
The criteria agree with previously published work, highlighting the occurrence of stall for tip-speed ratios $\lambda < 4$ and a transition from light to deep stall at a critical tip-speed ratio related to the vortex formation timescale.
To generalise our findings, we present a parametric map that predicts the expected stall regime based on the turbine's operating parameters, namely the tip-speed ratio $\lambda$ and chord-to-diameter ratio $c/D$.
These findings are valuable in the design of cross-flow and vertical-axis wind turbine geometries and operating conditions as the occurrence and degree of stall govern the expected performance of wind turbines.

%				--------	  			~		experimental setup		~					--------				%

\section{Experimental apparatus and methods}

%				--------	  			~		experimental setup		~					--------				%

\begin{figure*}[tb!]
	\centering
	\includegraphics[width=0.7\linewidth]{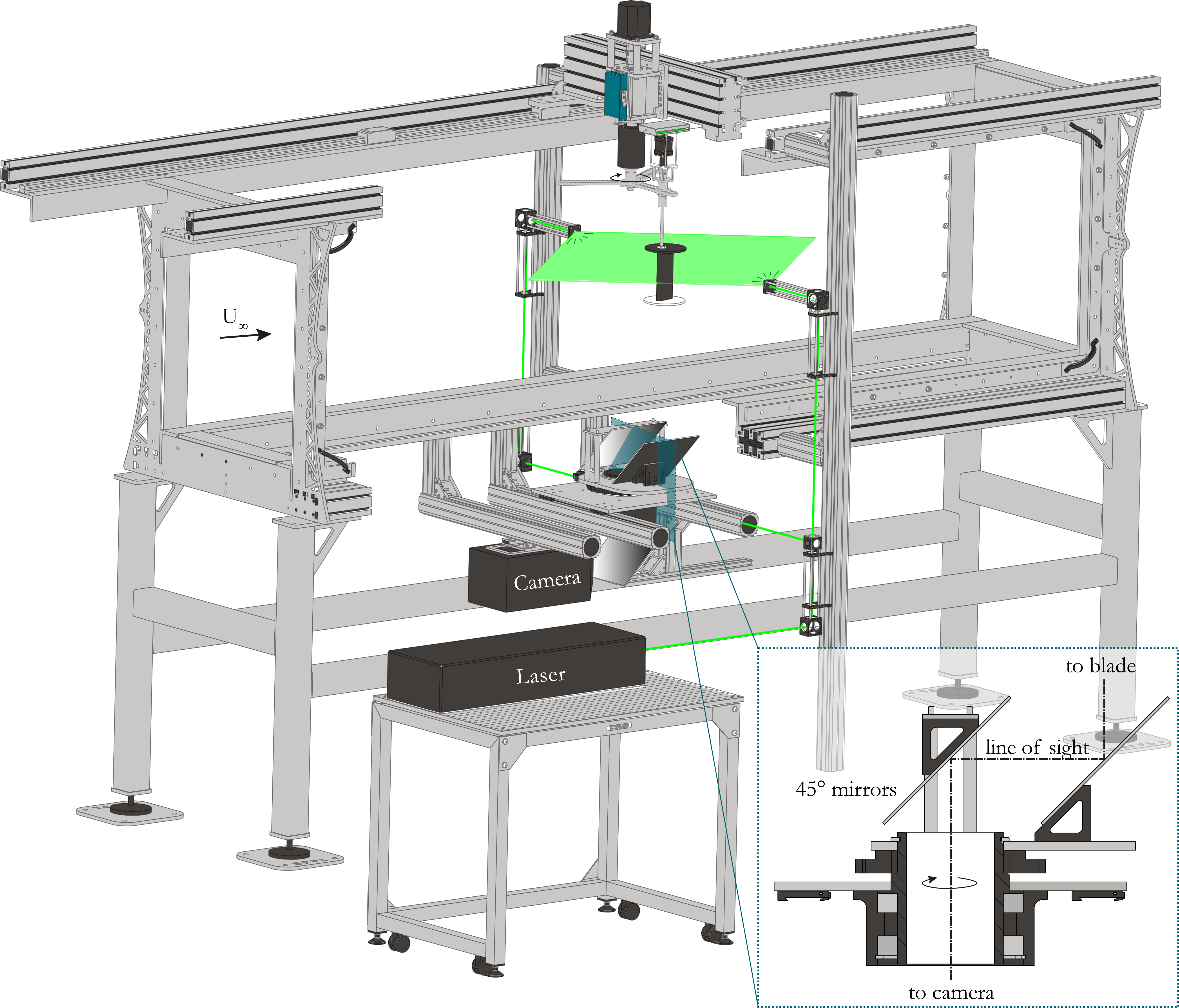}
	\caption{Schematic of the experimental set-up including the wind turbine model, the light sheet, the rotating mirror system and high-speed camera for particle image velocimetry. A section view of the rotating mirror system is also depicted in the bottom right corner.}
	\label{fig:expsetup}
\end{figure*}

Experiments were conducted in a recirculating water channel with a test section of  \SI{0.6x0.6x3}{\meter} and a maximum flow velocity of \SI{1}{\meter\per\second}.
The test section has transparent acrylic walls that are bound by a metallic frame, providing optimal optical access for velocity field measurements.
A representation of the full experimental apparatus is shown in \cref{fig:expsetup}.

\subsection{Vertical-axis wind turbine model}

A scaled-down model of a single-bladed H-type Darrieus wind turbine was mounted in the centre of the test section.
The turbine has variable diameter $D$ that was kept constant here at \SI{30}{\centi\meter}.
Up to three blades can be attached to the rotor arms through straight shafts that are held from the top.
Here, we used the single blade configuration to focus on the flow development around the blade in the absence of interference from the wakes of other blades.
The turbine blade itself was 3D printed using photosensitive polymer resin (Formlabs Form 2 stereolithography), sanded with very fine P180 grit paper and covered with black paint.
The blade has a NACA0018 profile with a span of $s=\SI{15}{\centi\meter}$ and a chord of $c=\SI{6}{\centi\meter}$, yielding a chord-to-diameter ratio of $c/D=\num{0.2}$.
The chord-to-diameter ratio influences the development of dynamic stall of a vertical-axis wind turbine and is analogous to the blades' reduced frequency \cite{Parker2017}.
High chord-to-diameter ratios lead to shorter rotation cycles in terms of the blades' convective time.
The non-dimensional convective time is typically calculated as the ratio of the distance traveled by an airfoil or a blade divided by its chord length.
It is a non-dimensional measure of time.
For our turbine blade, the distance traveled within a full rotation equals $\pi D$ and the convective time is proportional to the reciprocal of the chord-to-diameter ratio.
High chord-to-diameter ratios thus shorten the convective time available for flow development within a single rotation cycle of the turbine and increase the likelihood of blade-wake interaction, even for a single-bladed wind turbine interacting with its own wake in the downwind half of its rotation.
In this investigation, the chord-to-diameter ratio was kept constant at the higher end of the typical operating conditions of wind turbines \cite{Parker2017}.
The turbine's compact geometry allowed for a relatively small blockage ratio of \SI{12.5}{\percent}, based on the ratio of the blade's frontal swept area and the water channel's cross section.
The blade is held by a cantilevered shaft such that there is no central strut interference with the flow.
At low tip-speed ratio ($\lambda < 4$), the effective blockage is closer to \SI{2.5}{\percent}, which is the blockage ratio calculated based on the ratio of the blade area to the cross-sectional area.
Additionally, a \num{2.5} chord length distance to the water channel' side walls is also respected at all times.
Based on these observations, we consider the blockage and confinement effects small and no correction to force measurements are applied \cite{Parker2017,Ross2020}.
This assumption is supported by a numerical simulation of a single-bladed wind turbine with a \SI{32}{\percent} blockage ratio \cite{SimaoFerreira2009}.
The turbine model is driven by a NEMA 34 stepper motor with a \ang{0.05} resolution for the angular position.
The rotational frequency was kept constant at \SI{0.89}{\hertz}, yielding a constant chord-based Reynolds number of $\kindex{\Rey}{c} = (\rho \omega R c)/\mu = \num{50000}$, where $\rho$ is the density and $\mu$ the dynamic viscosity of water.
The turbine's rotational frequency was kept constant to maintain a constant chord-based Reynolds number throughout the experiments.
To investigate the role of the tip-speed ratio in the occurrence of dynamic stall, we systematically vary the water channel's incoming flow velocity from \SIrange{0.14}{0.70}{\meter\per\second} to obtain tip-speed ratios ranging from \numrange{1.2}{6}.

\subsection{Load measurements}

The blade shaft was instrumentalised to record unsteady aerodynamic loads acting on the blade using twenty strain gauges forming five full Wheatstone bridge channels.
The strain gauges are powered and their output signal is amplified using an instrumentation amplifier with precision voltage reference placed on a printed circuit board that is mounted directly on the rotor arm.
The load cell was calibrated in-situ using a fully orthogonal calibration rig and undergoing \num{1350} independent loading conditions.
A calibration matrix was obtained by performing a linear regression on the calibration measurements.
This matrix contained the \SI{95}{\percent} confidence interval of the coefficients allowing an estimation of the uncertainty related to numerous factors including response linearity, hysteresis, repeatability, and measurement error.
The uncertainty was found to be below \SI{5}{\percent} for the shear force and pitching moment components.
A full description of the calibration procedure can be found in \cref{sec:ap_lc}, including the loading conditions, calibration matrix calculation and error estimation.

For each experiment, the wind turbine model starts at rest with the blade facing the oncoming flow.
The turbine blade is accelerated to its prescribed rotational speed.
After reaching the target rotational speed, we wait for five full turbine rotations before starting the load recordings.
Aerodynamic forces acting on the turbine blade are recorded at \SI{1000}{\hertz} for \num{100} full turbine rotations, then the blade is brought to rest.
The forces presented in this paper are two shear forces applied at the blade's mid-span in the radial \kindex{F}{R} and azimuthal \kindex{F}{$\theta$} direction, and the pitching moment about the blade's quarter-chord \kindex{M}{1/4} (\cref{fig:vawtaero}).
The total force applied to the blade is computed by combining the two shear forces: $\kindex{F}{tot} = \sqrt{\kindex{F}{R}^2 + \kindex{F}{$\theta$}^2}$.
All force coefficients are non-dimensionalised by the blade chord $c$, the blade span $s$, and the blade velocity $\kindex{\U}{b} = \omega R$ such that:
\begin{equation*}
\kindex{C}{tot/R/$\theta$} = \frac{\kindex{F}{tot/R/$\theta$}}{0.5\rho \kindex{\U}{b}^2 sc}\quad.
\end{equation*}
The subscripts tot, R, or $\theta$ refer to the total force, the radial, or the tangential force component.
The centripetal force resulting from the turbine's rotation was experimentally measured by operating the wind turbine in air.
The added drag from the two splitter plates was measured and offset for all investigated tip-speed ratios by operating the wind turbine without the blade, where the two splitter plates where held by a small cylinder.
The influence of the centripetal force and the splitter plate are subtracted from the raw measurement data to isolate and compare the aerodynamic forces acting on the turbine blade.
A description of the measurement and modelling of the non-aerodynamic forces is included in \cref{sec:ap_offset}.
The presented force date was filtered using a second-order low-pass filter with the cut-off frequency at \SI{30}{\hertz}.
This frequency is multiple orders of magnitude larger than the pitching frequency and approximately \num{50} times larger than the expected post-stall vortex shedding frequency based on a chord-based Strouhal number of \num{0.2} \cite{Leishman1989}.

\subsection{Particle image velocimetry}

High-speed particle image velocimetry (PIV) was used to measure the flow field around the wind turbine blade.
A dual oscillator diode pumped ND:YLF laser ($\lambda = \SI{527}{\nano\meter}$) with a maximum pulse energy of \SI{30}{\milli\joule} and a beam splitter were used to create two laser sheets from opposite sides of the channel.
The light sheets were oriented horizontally at mid-span of the turbine blade (\cref{fig:expsetup}).
A high speed camera with a sensor size of \SI{1024x1024}{\pixel} (Photron Fastcam SA-X2) and a spinning mirror apparatus were installed below the channel to capture the flow around the blade.
The spinning mirror apparatus comprises of two rotating and one stationary mirror, all oriented onto a \ang{45} plane with respected to the horizontal plane.
The two moving mirrors rotate about the same axis of rotation and at the same frequency as the wind turbine.
One of the mirrors is placed at the same radius as the model blade, such that it keeps the blade in the field of view of the camera.
The spinning mirror apparatus allows us to measure the velocity field around the blade with a higher spatial resolution and without scarifying the temporal resolution.
The field of view is \SI{2.5x2.5}{c} centred around the blade.
The acquisition frequency is \SI{1000}{\hertz}.
The images were processed following standard procedures using a multi grid algorithm \cite{Raffel2007}.
The final window size was \SI{48x48}{\pixel} with an overlap of \SI{75}{\percent}.
This yields a grid spacing or physical resolution of $\SI{1.7}{\milli\meter} = 0.029c$.

\subsection{Phase-averaging}

Load measurements are obtained and phase-averaged over \num{100} wind turbine revolutions for all tip-speed ratio cases.
Phase-averaging involves splitting the phase-space into \num{540} bins that cover \ang{0.67} without overlap.
%The data from all \num{100} turbine rotations is allocated into the corresponding bin.
For each bin, we calculate the mean performance value and its standard deviation.
This method allows us to visualise the mean performance of the turbine blade at \num{540} phase positions and the corresponding cycle-to-cycle variations of the performance.
The number of bins was selected to be large enough such that sufficient data points lie in each bin, and small enough such to reduce the smoothening of the data.
The acquired data and bin size yielded converged average and first-order statistical metrics of the unsteady aerodynamic loads experienced by the blade.
For the tip-speed ratios $\lambda \in \{1.2, 1.5, 2, 2.5, 3\}$, we captured time-resolved PIV for \num{19} wind turbine revolutions.

%				--------	  			~			results 			~					--------				%

\section{Results}

%				--------	  			~			results 			~					--------				%

\begin{figure*}[p]
\centering
\includegraphics{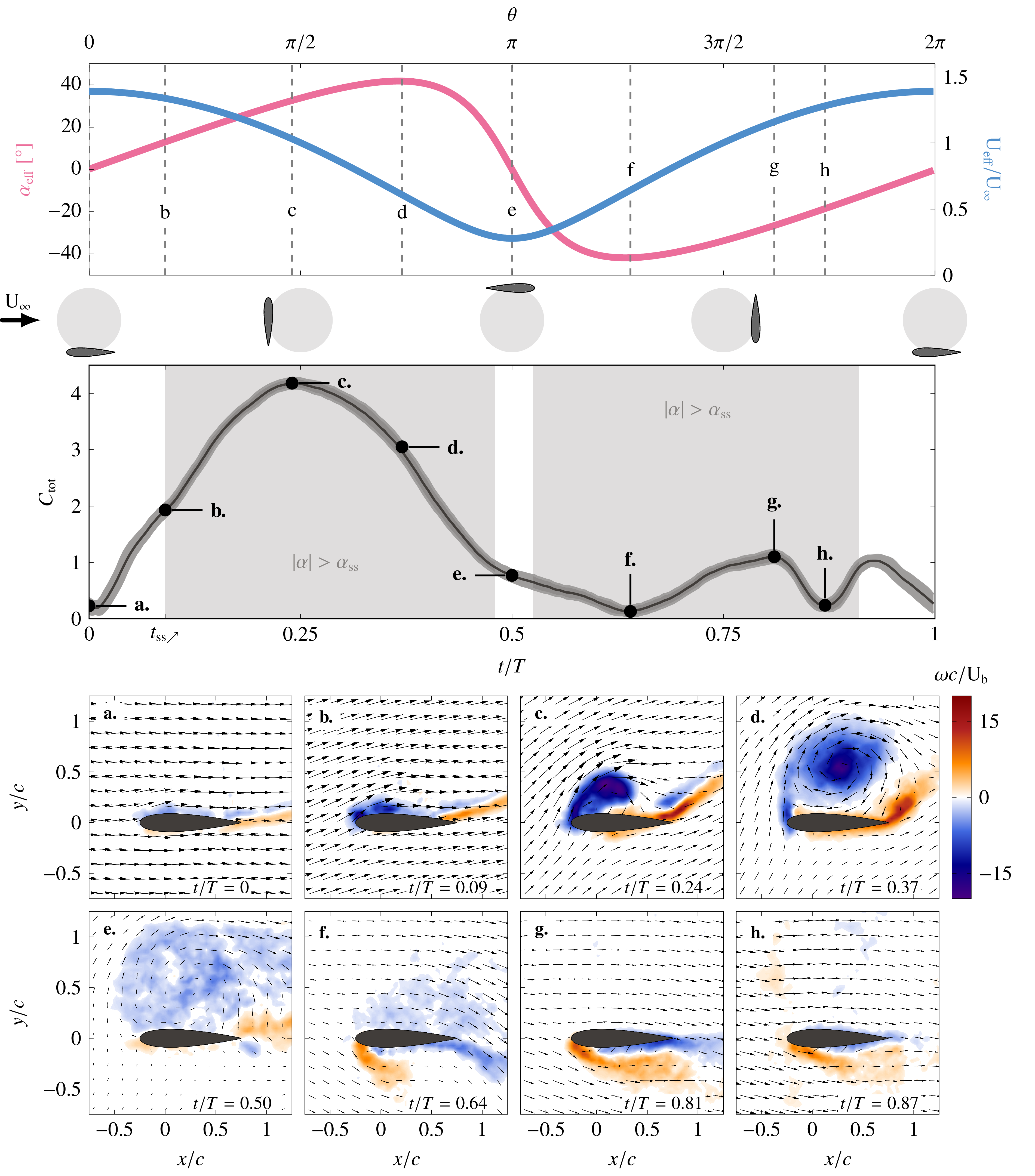}
\caption{Temporal evolution of the effective angle of attack $\kindex{\alpha}{eff}$, effective flow velocity $\kindex{\U}{eff}$, and total force coefficient $\kindex{C}{tot}$ experienced by the wind turbine blade operating at the tip-speed ratio $\lambda = 1.5$.
The blade's progression along its circular path relative to the incoming flow is depicted at the top of the figure.
Phase-averaged vorticity and velocity fields at eight points of interest \subf{a}-\subf{h}.
The flow fields are rotated in the frame of reference of the blades.
The vectors show the direction of the incoming flow and their relative lengths match the relative spatiotemporal differences in the velocity magnitude.}
\label{fig:Ctot1}
\end{figure*}

The temporal evolution of the effective angle of attack $\kindex{\alpha}{eff}$, effective flow velocity $\kindex{\U}{eff}$, and the phase-averaged total force coefficient $\kindex{C}{tot}$ experienced by the wind turbine blade operating at the tip-speed ratio $\lambda = 1.5$ are presented in \cref{fig:Ctot1}.
The blade's progression along its circular path relative to the incoming flow during one turbine rotation $T$ is depicted at the top of the figure.
%\subf{a} beginning of the rotation,
%\subf{b} blade exceeds its critical stall angle,
%\subf{c} first maximum load response,
%\subf{d} maximum effective angle of attack - dynamic stall vortex pinch-off,
%\subf{e} half-way through the blade's rotation,
%\subf{f} first minimum load response,
%\subf{g} second maximum load response,
%\subf{h} second minimum load response.
The phase-averaged vorticity and velocity fields at eight selected points of interest \subf{a}-\subf{h} are shown at the bottom of \cref{fig:Ctot1}.
The velocity field is captured in the blade's frame of reference, so it shows the effective flow conditions acting on the blade.

At the beginning of the blade's revolution ($t/T=0$), the blade sees the incoming flow under a zero angle of attack and the flow around the blade is attached (\cref{fig:Ctot1}\subf{a}).
The small total force experience by the blade is attributed to unsteady aerodynamic effects caused by the high change in the blades' angle of attack or pitch rate at $t/T = 0$ \cite{Bensason2021}.
During the initial part of the rotation, the angle of attack is positive and increases gradually until it reaches a maximum value of $\kindex{\alpha}{eff}=\num{42}$ at $t/T=0.37$.
At $t/T=0.09$, the angle of attack exceeds the critical static stall angle of $\kindex{\alpha}{ss}=\ang{13.1}$ that was determined for the turbine's blade at $Re = 50k$ using the procedure described in \cite{LeFouest2021}.
Shortly after the blade's angle of attack has exceeded its static stall value, vorticity starts to accumulate near the leading edge (\cref{fig:Ctot1}\subf{b}) and forms a coherent leading edge or dynamic stall vortex (\cref{fig:Ctot1}\subf{c}).
During the formation of the dynamic stall vortex, the total force coefficient increases approximately linearly with time from $\kindex{C}{tot} = 0.5$ to $\kindex{C}{tot} \approx 4$ within a quarter of the rotation period.
This maximum value of the total force coefficient is well beyond the values that we observe for this airfoil under static conditions.
This lift overshoot is a characteristic feature of dynamic stall.
The decrease of the total force coefficient for $t/T>0.24$ is linked to the lift-off of the dynamic stall vortex away from the airfoil's surface and the emergence of positive or opposite-signed vorticity between the vortex and the surface.
After lift-off, the dynamic stall vortex continues to grow and reaches a size that is of the order of the blade's chord length (\cref{fig:Ctot1}\subf{d}).
Opposite sign vorticity has now spread across the entire chord length and leads to a loss of suction on the airfoil's surface and a drop in the total force coefficient.
The total force coefficient drops even faster when the angle of attack starts to decrease for $t/T>0.38$.
During this part of the cycle ($0.38<t/T<0.5$), the large scale dynamic stall vortex loses coherence and the vorticity dissipates (\cref{fig:Ctot1}\subf{d,e}).
When the blade enters the second half of its rotation, the blade is surrounded by a heavily separated flow and the remnants of the dynamic stall vortex that is convected downstream (\cref{fig:Ctot1}\subf{e}).
The blade now also moves in the downstream direction and continues to interact with its own wake until $t/T\approx 0.6$.
This contributes to a further drop in the force coefficient to almost zero.
The local minimum of the force coefficient coincides with the moment at which the effective angle of attack has reached its maximum negative value.
For the remainder of the cycle, the suction side of the blade is now on the outside of the circular trajectory where positive vorticity accumulates (\cref{fig:Ctot1}\subf{f}).
This vorticity accumulates into a vortex that is stretched along the chord length and gives raise a local force maximum $\kindex{C}{tot} = 1.1$ at $t/T = 0.8$ (\cref{fig:Ctot1}\subf{g}).
This vortex does not have much time to grow and is convected downstream when the absolute value of the effective angle of attack decreases and the flow reattaches along both sides of the blade (\cref{fig:Ctot1}\subf{h,a}).

\begin{figure*}[p]
\centering
\includegraphics{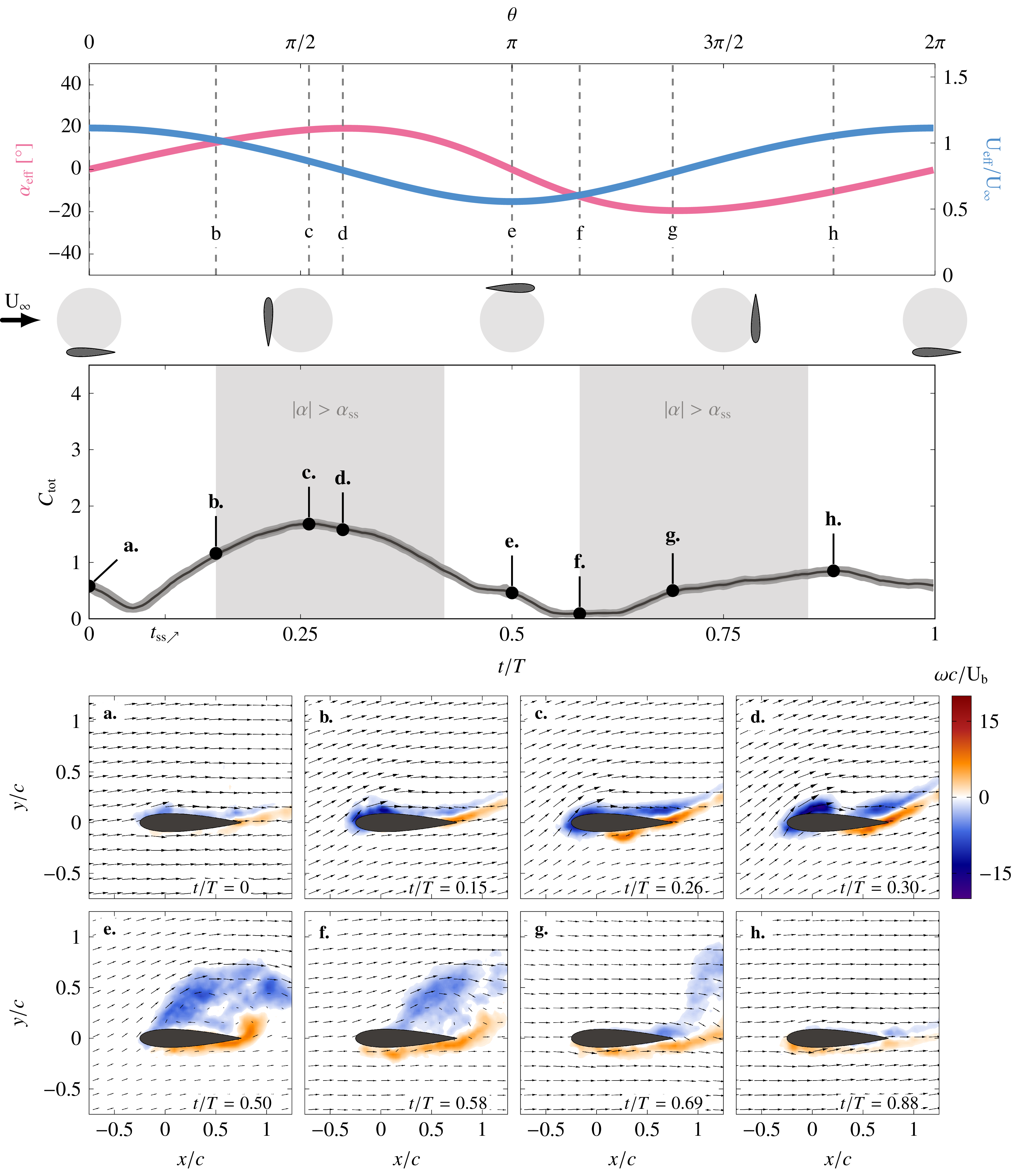}
\caption{Temporal evolution of the effective angle of attack $\kindex{\alpha}{eff}$, effective flow velocity $\kindex{\U}{eff}$, and total force coefficient $\kindex{C}{tot}$ experienced by the wind turbine blade operating at the tip-speed ratio $\lambda = 3.0$.
The blade's progression along its circular path relative to the incoming flow is depicted at the top of the figure.
Phase-averaged vorticity and velocity fields at eight points of interest \subf{a}-\subf{h}.
The flow fields are rotated in the frame of reference of the blades.
The vectors show the direction of the incoming flow and their relative lengths match the relative spatiotemporal differences in the velocity magnitude.
}
\label{fig:Ctot2}
\end{figure*}

The blade kinematics $\kindex{\alpha}{eff}$ and $\kindex{\U}{eff}$, phase-averaged total force coefficient $\kindex{C}{tot}$ and flow fields for the tip-speed ratio $\lambda = 3.0$ are presented in \cref{fig:Ctot2} for comparison.
The main differences with the previously described development at $\lambda = 1.5$ are a lower total force magnitude, the smaller size of the dynamic stall vortex and a reduction in post-stall fluctuations.
The total force magnitude in the upwind half of the blade's rotation ($0 \leq t/T < 0.5$) is \num{2.5} times smaller for $\lambda = 3.0$ compared to $\lambda = 1.5$ due to the lower pitching and surging amplitude experienced by the blade operating at $\lambda = 3.0$.
The dynamic stall vortex that forms during the upwind half of the cycle at $\lambda = 3.0$ is much smaller and less coherent than the vortex formed at $\lambda = 1.5$.
For $\lambda = 3.0$, the dynamic stall vortex does not cover more than half of the chord length (\cref{fig:Ctot2}a-d).
The portion of the cycle that is influences by the interaction of the blade with the shed dynamic stall vortex is significantly reduced and the flow reattaches early in the downwind half of the rotation (\cref{fig:Ctot2}f-h).
In general, we observed less load fluctuations and a more symmetric load response for the first and second half of the rotation for higher tip-speed ratios.
When the tip-speed ratio $\lambda = \dfrac{\omega R}{\Uinf}$ decreases towards 1, the blade's effective flow condition see larger periodic oscillations, leading to more dominant dynamic stall events, greater force production, and larger post-stall load transient.

\begin{figure}[b!]
\centering
\includegraphics{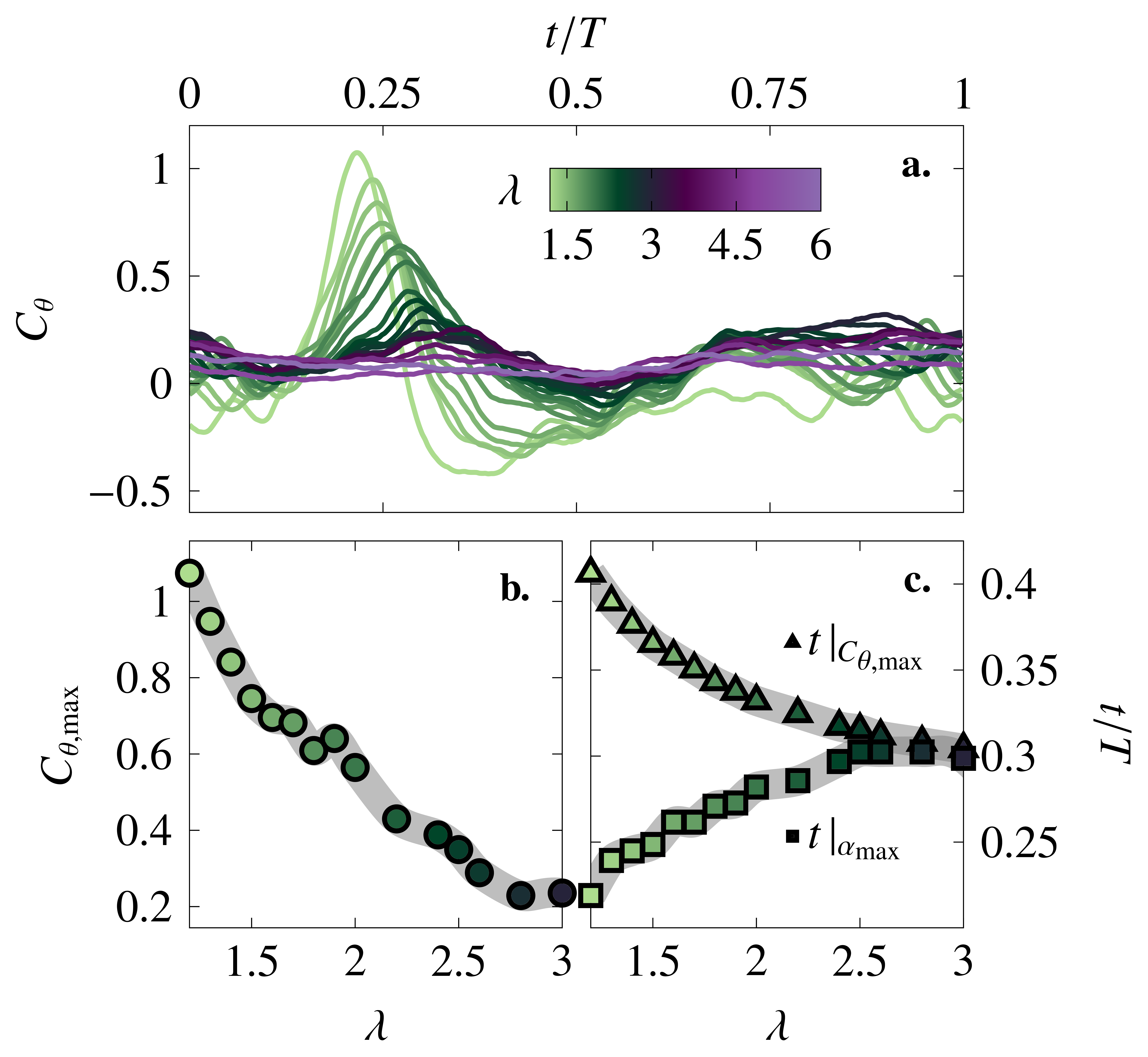}
\caption{\subf{a} Temporal evolution of the tangential force coefficient $\kindex{C}{$\theta$}$ experienced by the wind turbine blade operating at tip-speed ratios ranging from \numrange{1.2}{6}.
\subf{b} Maximum values of the tangential force coefficient as a function of the tip-speed ratio.
\subf{c} Temporal occurrence of the tangential force maximum $t|_{\kindex{C}{$\theta$,max}}$ and temporal occurrence of the maximum angle of attack $t|_{\kindex{C}{$\alpha$,max}}$ as a function of the tip-speed ratio.
We focus on tip speed ratio $\lambda \in [1.2 - 3]$ as they have a clearly defined maximum that can readily be identified. %\Bigr
}
\label{fig:Ct_all}
\end{figure}

The power production of vertical axis wind turbines is directly proportional to the force component tangential to the blade.
The temporal evolutions of the tangential force coefficient $\kindex{C}{$\theta$}$ experienced by the wind turbine blade operating at tip-speed ratios ranging from \numrange{1.2}{6} are presented in \cref{fig:Ct_all}\subf{a}.
At a high tip-speed ratios ($\lambda\geq\num{4.5}$), the blade experiences almost no tangential force during the entire rotation.
For mid and low tip-speed ratios ($\lambda<\num{4.5}$), the tangential force coefficient features a distinct peak in the upwind half that increases in magnitude (\cref{fig:Ct_all}\subf{b}) with decreasing tip-speed ratio (\cref{fig:Ct_all}\subf{b}).
The maximum tangential force coefficient increases from \num{0.24} at $\lambda = 3$ to \num{0.35} at $\lambda = 2.5$.
In this range of tip-speed ratios, positive tangential force is mainly observed in the second part of the upwind part of the rotation and is attributed to the presence of a dynamic stall vortex (\cref{fig:Ctot2}\subf{d}).
If we further reduce the tip-speed ratio, a stronger dynamic stall vortex is observed, which starts to form earlier in the cycle (\cref{fig:Ctot1}\subf{c} vs \cref{fig:Ctot2}\subf{d}).
This results in a strong increase in the maximum tangential force from \num{0.35} at $\lambda = 2.5$ to \num{1.2} at $\lambda = 1.2$.
The peak tangential force occurs earlier in the cycle the lower the tip-speed ratio.
For $\lambda < 2.5$, the peak is followed by a significant drop and excursion in negative values (\cref{fig:Ct_all}\subf{a}).
Negative tangential force is equivalent to negative instantaneous torque generated by the blade and is undesirable for a vertical-axis wind turbine.

The variation in the maximum tangential force coefficient with tip-speed ratio follows a bi-linear evolution with a transition point around $\lambda = 2.5$.
The transition point corresponds to the conditions for which the maximum tangential force coefficient occurs at the maximum angle of attack and marks a fundamental shift in the unsteady blade load response (\cref{fig:Ct_all}\subf{d}).
For $\lambda > 2.5$, the tangential force evolution follows the blade kinematics and the force maximum is governed by $\kindex{\alpha}{max}$.
For $\lambda < 2.5$, the tangential force evolution does not purely follow the blade kinematics anymore and is strongly affected by the occurrence of dynamic stall.
This leads to a strong increase in the tangential force overshoot with decreasing tip-speed ratio but also a increase in the force fluctuations following the separation of the dynamic stall vortex.

\begin{figure*}[tb]
\centering
\includegraphics{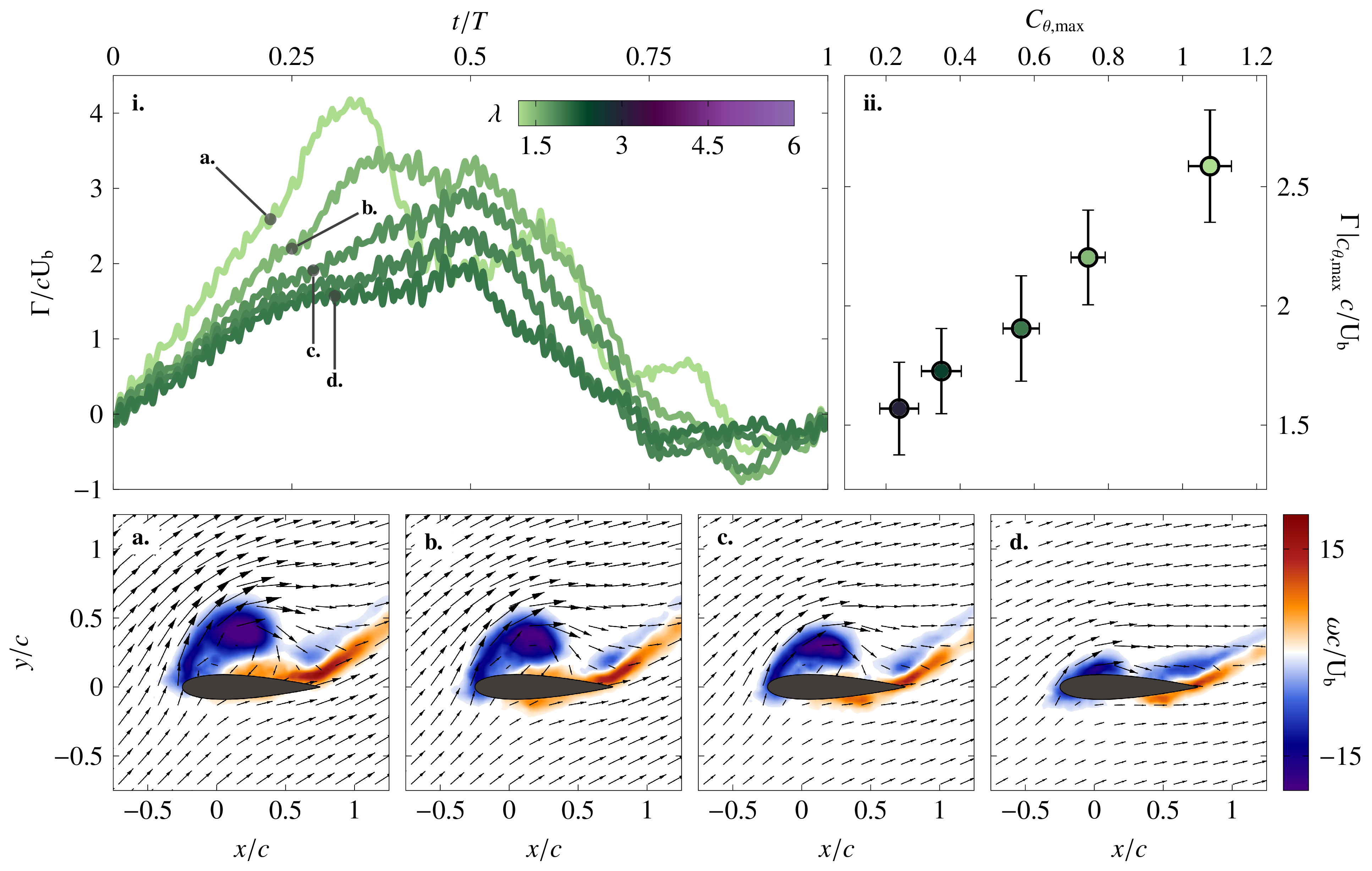}
\caption{\subf{a} normalised circulation around the turbine blade against normalised time for tip-speed ratios  $\lambda = {1.2, 1.5, 2.0, 2.5, 3.0}$. absolute normalised circulation taken at the moment when the maximum tangential force is reached ($\Gamma\rvert_{\kindex{C}{$\theta$,max}} / c\kindex{U}{b}$) against \subf{b} tip-speed ratio and \subf{iii} maximum tangential force coefficient $\kindex{C}{$\theta$, max}$.}
\label{fig:Circ}
\end{figure*}

To further evaluate the influence of dynamic stall on the tangential force coefficient, we analyse the temporal evolution of the circulation around the blade for five tip-speed ratio cases $\lambda = {1.2, 1.5, 2.0, 2.5, 3.0}$ in \cref{fig:Circ}.
The circulation was computed by integrating all the phase-average vorticity in the entire field of view surrounding the blade. %\todo{all vorticity or only negative signed vorticity?}
The circulation values were non-dimensionalised by the blade velocity and chord length.
The absolute value of the normalised circulation is zero at the start of the upwind rotation and starts increasing approximately linearly in time with increasing rate for decreasing tip-speed ratios.
The end of the linear increase region corresponds to the moment when the maximum tangential force is reached.
At this point, the flow features a coherent leading-edge dynamic stall vortex which increases in size and strength with decreasing tip-speed ratio \cref{fig:Circ}\subf{a-d}.
The absolute normalised circulation taken at the moment when the maximum tangential force coefficient is reached $\Gamma / c\kindex{U}{b} @ \kindex{C}{$\theta$,max}$ is presented versus the maximum tangential force coefficient in \cref{fig:Circ}\subf{ii}.
A higher maximum tangential force coefficient is directly related to a stronger dynamic stall vortex  \cref{fig:Circ}\subf{ii}.
This stronger dynamic stall vortex has also formed quicker than at lower tip speed ratios.
The snapshots in \cref{fig:Circ}\subf{a-d} present the vortex at the moment in the cycle when the maximum tangential force is reached which occurs earlier for lower tip-speed ratios.
From left to right, the time at which the snapshots are taken increases but we seem to be looking at the formation process of the vortex in reversed time.
This highlights the importance of various time scales that play a role in the flow and force response on vertical-axis wind turbine blades.

After the maximum in tangential force coefficient is reached, the circulation associated with the dynamic stall vortex continues to increase.
The blade is now in the second half of the upwind rotation, it is moving along the direction of the incoming flow and experiences a decreasing effective velocity.
As a result, the vorticity feeding rate of the dynamic stall vortex decreases and the vortex lifts-off from the surface.
Vortex lift-off occurs when the vortex has grown strong enough such that secondary vortices of opposite-signed vorticity emerge between the vortex and the blade's surface (\cref{fig:Circ}\subf{a}) or when the effective inflow velocity decreases and drops below the self-induced velocity of the vortex \cite{Rival:2014bf, Kissing2020}.
Even though the dynamic stall vortex and the circulation in the field of view continues to increase for $0.25\leq t/T<0.4$, the tangential force coefficient drops as the vortex is no longer bound to the blade.
The drop in the tangential force is followed by undesirable load transients that increase structural stress, lead to potentially vibrations, and sometimes even turbine failure.
Here, we will now propose and present two methods to quantify these undesirable load fluctuations.

\begin{figure}[tb]
\centering
\includegraphics{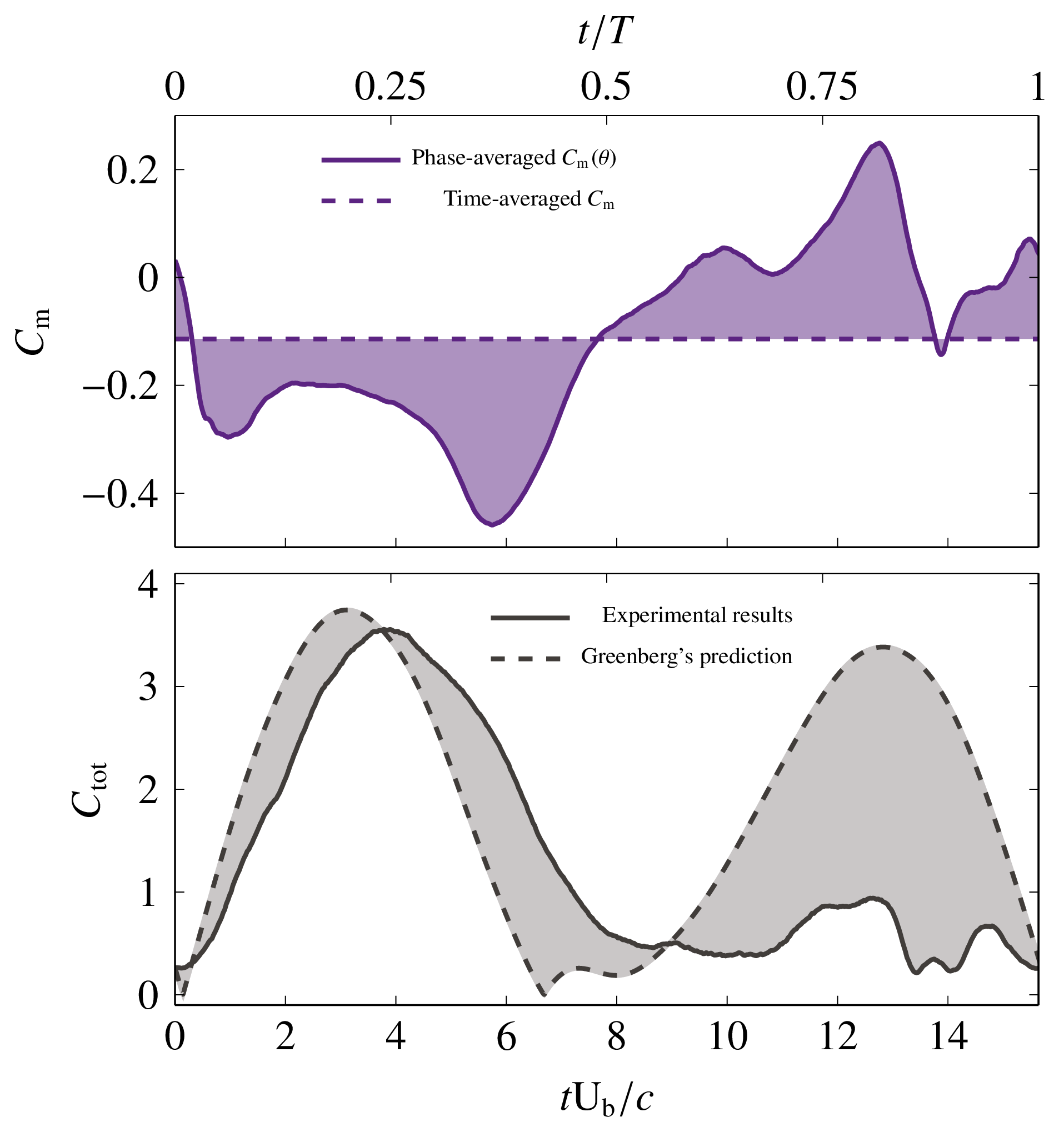}
\caption{
Proposed quantities to assess the magnitude of stall induced load transients based on \subf{a} the phase-averaged pitching moment standard deviation;
\subf{b} the different between the phase-averaged total force coefficient $\kindex{C}{tot}$ and the inviscid solution obtained using an adapted Greenberg model from \cite{Bensason2021}.
Examples include data for $\lambda = 1.5$ and $\lambda = 1.7$.
}
\label{fig:quantLT}
\end{figure}

The first method consists of quantifying the standard deviation of the phase-averaged pitching moment acting on the blade (\cref{fig:quantLT}\subf{a}).
By definition, the pitching moment about the airfoil's aerodynamic centre is independent of the angle of attack.
For a NACA0018, the aerodynamic centre is close to the quarter-chord.
An increase in the standard deviation of the measured pitching moment serves as a measure of the extent of flow separation occurring on the blade.
The pitching moment is also directly related to the torsional stress applied to the blade shaft.
Significant transients of this metric are undesirable from a turbine structural integrity perspective.

The second method consists of computing the mean difference between the total phase-averaged force experienced by the blade and the total force computed using Greenberg's model $\Delta \kindex{C}{total}$ (\cref{fig:quantLT}\subf{b}).
Greenberg's model is an inviscid flow model that is used to predict the unsteady aerodynamic loads acting on a blade placed in a free-stream and undergoing any combination of oscillatory pitching, surging, and heaving \cite{Greenberg1947}.
Greenberg's model was adapted here to accommodate the kinematics experienced by a wind turbine blade following the approach described by \cite{Bensason2021}.
The Greenberg model does not account for vorticity-induced loads related to flow separation.
The model also does not account for the fact that the blade follows a circular path which gives rise to apparent camber effects \cite{Bianchini:2016ba}.
Large differences between the experimental loads and the inviscid solution indicate an increased influence of flow separation and blade-wake-interactions.

\begin{figure}[b!]
\centering
\includegraphics{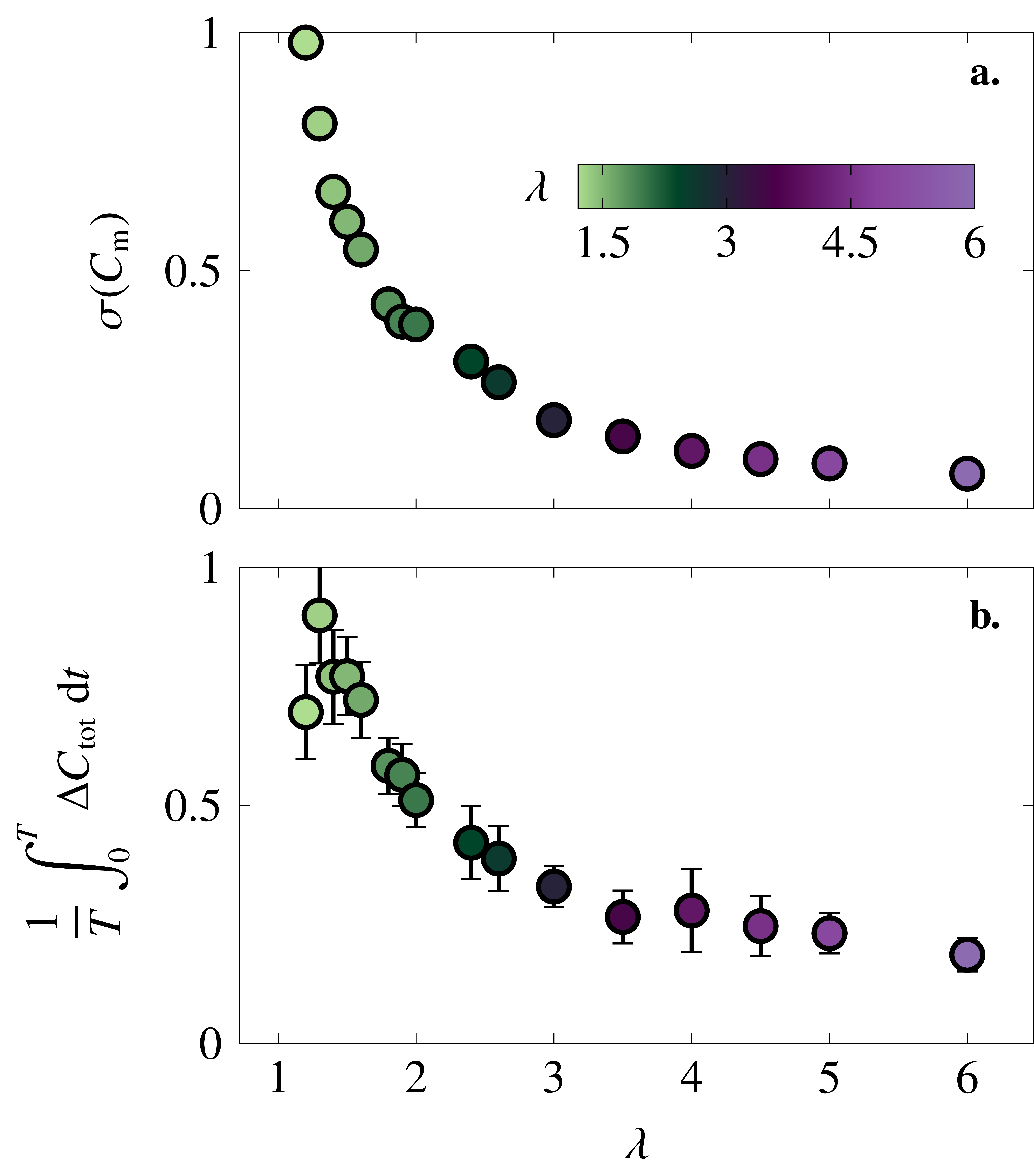}
\caption{\subf{a} Phase-averaged pitching moment standard deviation and \subf{b} mean difference between the total force coefficient from experiments and the total force coefficient computed using the adapted Greenberg model from \cite{Bensason2021}.}
\label{fig:fluct}
\end{figure}

The phase-averaged pitching moment standard deviation and the mean difference in total force coefficient between experimental and theoretical results are presented in \cref{fig:fluct}.
Both metrics are normalised by the maximum value achieved over this range of tip-speed ratios to facilitate comparison.
The load fluctuations related to flow separation decrease with increasing tip-speed ratio following a power law decay.
The phase-averaged pitching moment standard deviation is largest for the lowest tip-speed ratio ($\lambda=1.2$).
The pitching moment standard deviation decreases from \numrange{0.90}{0.15} for the tip-speed ratio increasing from \numrange{1.2}{3.0}.
The difference between the experimentally measured total load and the inviscid solution follows a similar trend as the pitching moment standard deviation.
%, except for two relevant differences.
%The lowest tip-speed ratio does not correspond to the highest fluctuation levels  (\cref{fig:fluct}b).
%The dynamic stall load transients in the downwind half of the turbine rotation ($0.5 \leq t/T < 1$) coincidentally reduce the difference with the second total force maximum from the inviscid solution, related to blade kinematics.
%Secondly, the total force difference suggests an \SI{83}{\percent} reduction in fluctuation intensity between tip-speed ratio \num{1} and \num{6}, against \SI{93}{\percent} for the pitching moment standard deviation.
Both metrics are quick and easy to compute from instantaneous load measurements which makes them suitable metrics to use for future applications of flow control and closed loop performance optimisation.

\begin{figure*}[tb]
\centering
\includegraphics{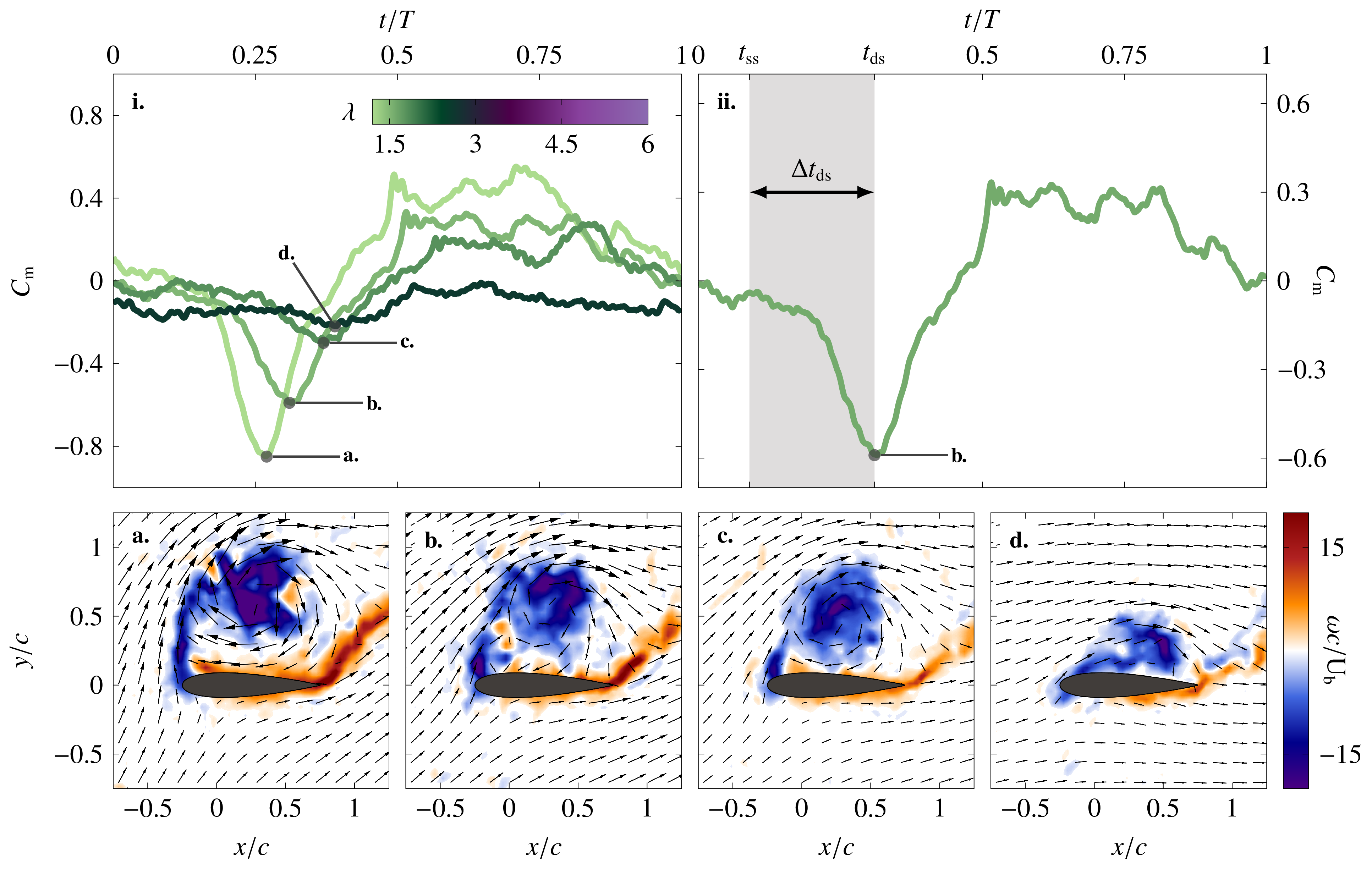}
\caption{\subf{i} Temporal evolution of the pitching moment coefficient for one blade rotation at tip-speed ratios $\lambda \in \{1.2, 1.5, 2, 3\}$;
\subf{ii} illustration of the definition and calculation of the dynamic stall delay;
\subf{a-d} instantaneous velocity and vorticity snapshots at vortex pinch-off, corresponding to the minimum pitching moment coefficient, for tip-speed ratios of \subf{a} $\lambda=1.2$, \subf{b} $\lambda=1.5$, \subf{c} $\lambda=2$, \subf{d} $\lambda=3$.}
\label{fig:Cm}
\end{figure*}

\begin{figure}[tb]
\centering
\includegraphics{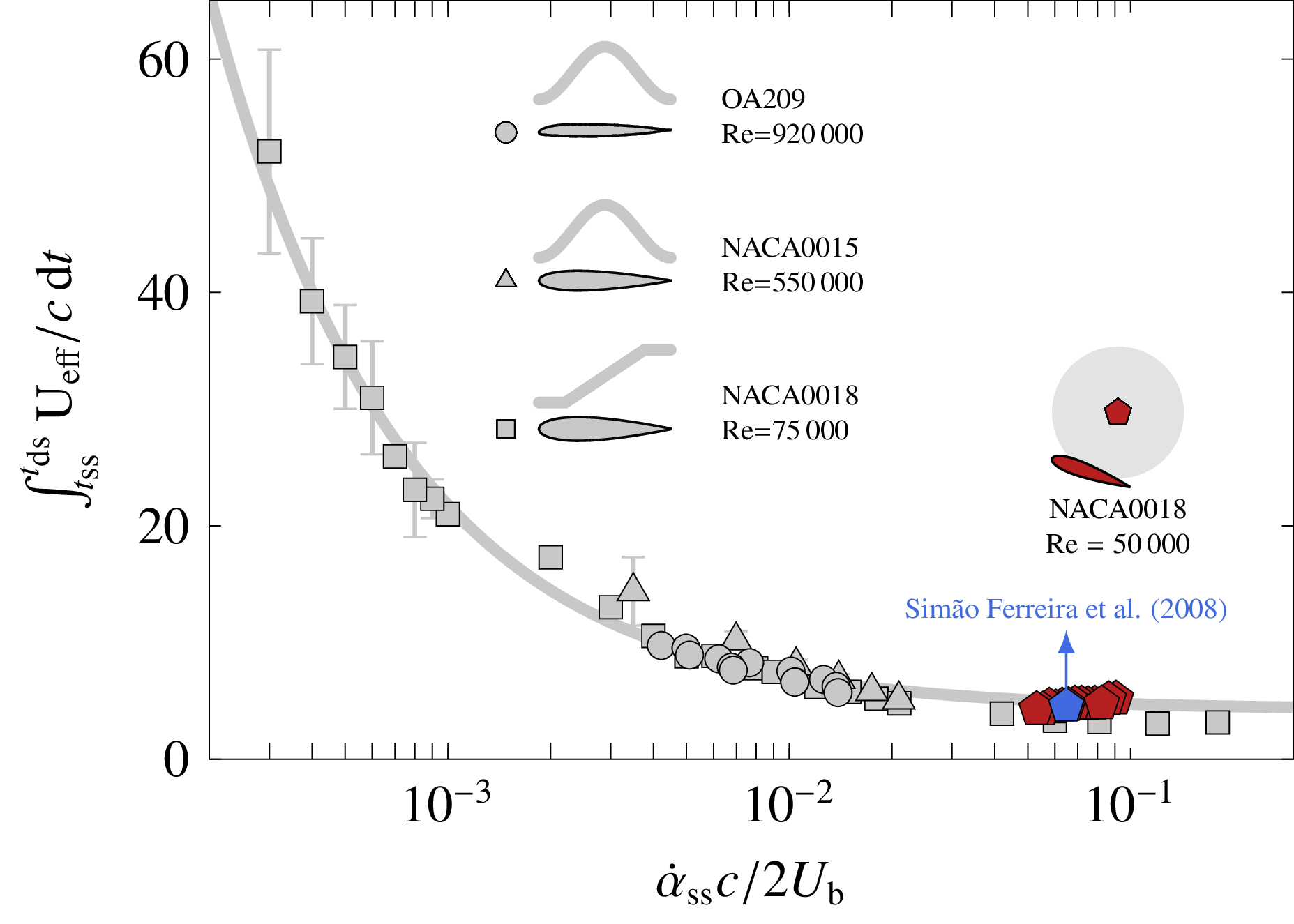}
\caption{Dynamic stall delay $\Delta \kindex{t}{ds}$ as a function of the reduced frequency for the vertical-axis wind turbine operating at $\lambda \in [1.2 - 4]$.
The results for the wind turbine blade are compared to those obtained for three other dynamic stall experiments: linear ramp-up motions of a NACA0015 at $Re = $\num{7.5e4} \cite{LeFouest2021}, a sinusoidally pitching NACA0015 at $Re = $\num{5.5e5} \cite{He2020}, a sinusoidally pitching 0A209 at $Re = $\num{9.2e5} \cite{Mulleners2012}, and one data point extracted from the single-bladed vertical-axis wind turbine data by \cite{SimaoFerreira2009}.
}
\label{fig:dsdelay}
\end{figure}

Based on the temporal evolution of the torque production, flow structures, and load transients, we can distinguish three operating regimes for vertical-axis wind turbines: a regime where no stall occurs, and regimes that are characterised by light or deep dynamic stall.
A parametric dynamic stall map indicating which regime a wind turbine is expected to operate in as a function of the tip-speed ratio and the chord-to-diameter ratio is desirable.
To determine a robust criterion allowing for the classification of the different tip-speed ratios, we revert to the pitching-moment coefficient around the blade's quarter-chord.
The temporal evolutions of the pitching moment coefficient during one turbine revolution at tip-speed ratios $\lambda \in \{1.2, 1.5, 2, 3\}$ are presented in \cref{fig:Cm}\subf{i}.
For all tip-speed ratios, we observe a minimum or negative peak pitching moment in the upwind part of the rotation.
When the dynamic stall vortex forms and grows, it increases the suction at the leading edge, leading to a nose-down, i.e. negative, pitching moment (\cref{fig:Cm}\subf{i}).
Once the feeding shear layer of the dynamic stall vortex is interrupted by opposite-signed vorticity close to the surface, the vortex is considered to be separated.
The separation or pinch-off of the primary stall vortex marks the onset of dynamic stall and is followed by the recovery of the pitching moment (\cref{fig:Cm}\subf{a-d}).
The minimum pitching moment coincides here with the leading-edge vortex separation and can be used as a reference to calculate the dynamic stall delay.

\begin{figure}[b!]
\centering
\includegraphics{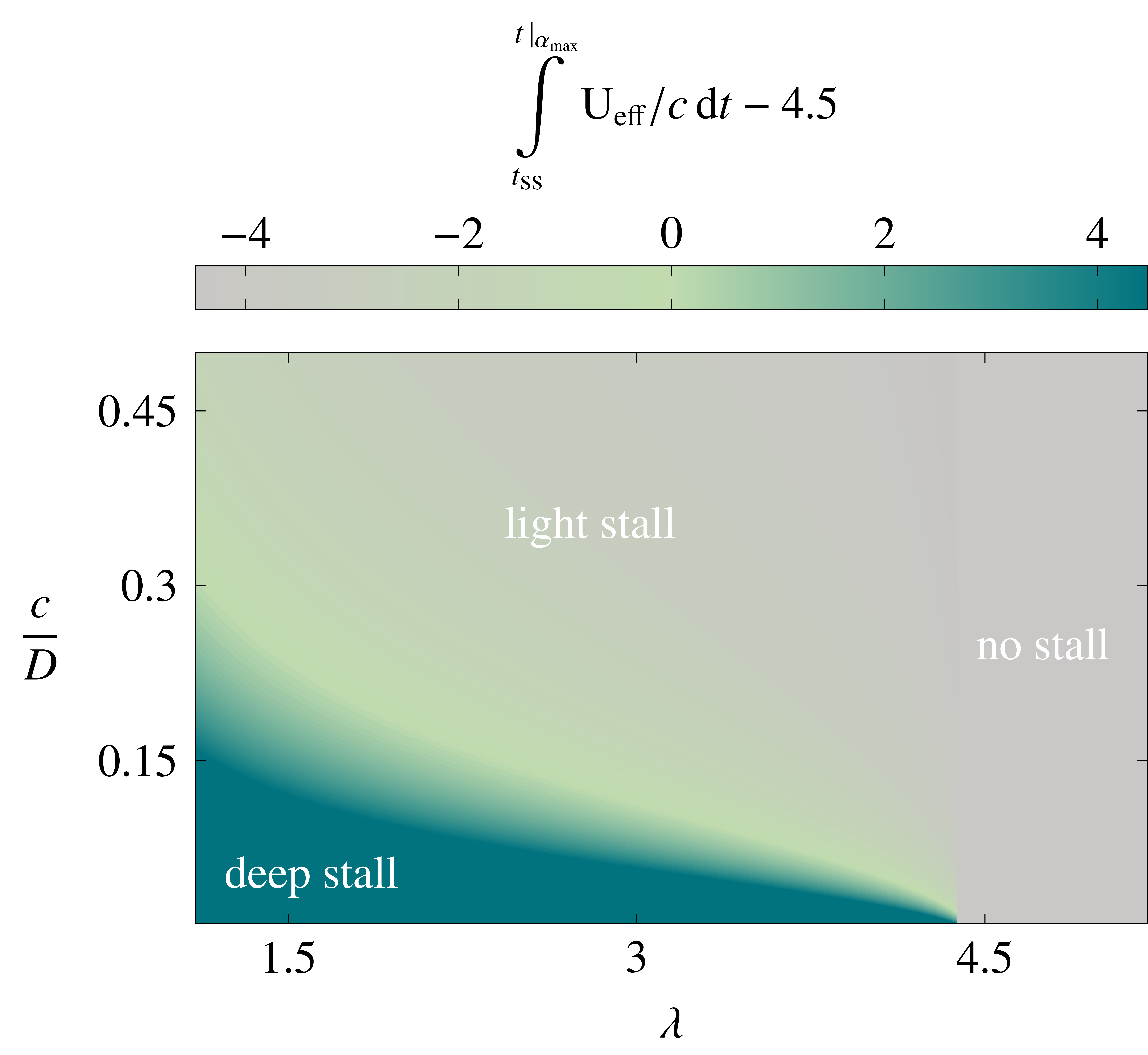}
\caption{Theoretical parametric dynamic stall regime map based on the timescales of the turbine blade kinematics and the characteristic timescales of dynamic stall development.
The difference between the available convective time for stall development and the expected dynamic stall delay for a given combination of the chord-to-diameter ratio $c/D$ and tip-speed ratio $\lambda$ is colour-coded and highlight the deep and light dynamic stall regimes.
The no-stall regime indicates operating conditions for which the blade does not exceed its static stall. }
\label{fig:DSmap}
\end{figure}

The dynamic stall delay is an important timescale for modelling the dynamic stall load response \cite{Ayancik2022}.
The stall delay is defined as the time interval between the moment when the blade exceeds its critical static stall angle of attack \kindex{t}{ss} and the onset of dynamic stall \kindex{t}{ds} (\cref{fig:Cm}\subf{ii}).
The non-dimensionalised dynamic stall delay follows a universal behaviour governed by the reduced pitch rate, which is largely independent from airfoil shape, motion kinematics, or Reynolds number \cite{LeFouest2021, Ayancik2022}.
We computed the dynamic stall delay for all wind turbine experiments where the blade exceeds its static stall angle, which is for tip-speed ratios $\lambda \in [1.2 - 4]$.
The effective velocity acting on the turbine blade changes in time and we determine the non-dimensional dynamic stall delay in terms of the airfoil's convective time scale $\int_{\kindex{t}{ss}}^{\kindex{t}{ds}} \kindex{U}{eff}/c\, \textrm{d} t$, as shown in \cite{Dunne2016a}.
We considered the pitch rate acting on the blade when it exceeds its static stall angle $\kindex{\dot{\alpha}}{ss}$ to determine the reduced frequency, such that $k = \kindex{\dot{\alpha}}{ss} c/2 \kindex{\U}{b}$.
The wind turbine dynamic stall delay is presented as a function of the reduced frequency in \cref{fig:dsdelay}.
The wind turbine results are compared to the general behaviour observed for four other dynamic stall experiments: linear ramp-up motions of a NACA0018 at $Re = $\num{7.5e4} \cite{LeFouest2021}, a sinusoidally pitching NACA0015 at $Re = $\num{5.5e5} \cite{He2020}, a sinusoidally pitching 0A209 at $Re = $\num{9.2e5} \cite{Mulleners2012}, and one data point extracted from the single-bladed vertical-axis wind turbine data by Simão Ferreira et al.~\cite{SimaoFerreira2009}.
The wind turbine dynamic stall delay falls on the same power law decay as the result of the non-rotating dynamic stall data and reaches the same values as the rotating data from \cite{SimaoFerreira2009}.
The agreement of vertical-axis wind turbine dynamic stall delay with different kinematics, airfoils, and Reynolds number suggests that the dynamic stall delay universality can be extended to pitching kinematics that include a surging component and a non-linear path.
Furthermore, the occurrence of deep dynamic stall on a vertical-axis wind turbine can be predicted based on its operating parameters, namely the tip-speed ratio $\lambda$ and chord-to-diameter ratio $c/D$.
Deep dynamic stall will occur when the dynamic stall onset occurs before the maximum effective angle of attack is reached \cite{Mulleners2012}.
By comparing the expected dynamic stall delay with the actual time interval available for dynamic stall to occur for given, predefined motion kinematics, we can create the stall regime map in \cref{fig:DSmap}.
For vertical-axis wind turbines, the reduced frequency is generally high enough, such that the dynamic stall delay varies minorly with the reduced pitch rate ($k>0.04$).
Vertical-axis wind turbine kinematics that yield a time interval between the moment the blade exceeds its static stall angle and when the moment it reaches its maximum effective angle that is longer than \num{4.5} convective ($\int_{\kindex{t}{ss}}^{\kindex{t}{ds}} \kindex{U}{eff}/c\, \textrm{d} t>4.5$) are expected to experience deep dynamic stall.

A theoretical parametric dynamic stall regime map is presented in \cref{fig:DSmap}.
This map was computed based on the theoretical timescales of turbine blade kinematics obtained for a wide range of tip-speed ratios $\lambda$ and chord-to-diameter $c/D$.
For each operating condition, we compute the available time for vortex formation as the time difference between the moment when the blade exceeds its critical static stall angle of attack \kindex{t}{ss} and the moment the maximum effective angle of attack is reached $t\,|_{\kindex{\alpha}{\tiny max}}$.
We calculate an excess convective time beyond the minimum required dynamic stall delay or vortex formation convective time as $\int_{\kindex{t}{ss}}^
{t\,|_{\kindex{\alpha}{\tiny max}}}\kindex{U}{eff}/c \, \textrm{d} t$ with the average dynamic stall delay of \num{4.5} required for highly unsteady motions (\cref{fig:dsdelay}).
Negative values of the excess time indicate operating conditions where the dynamic stall vortex formation is interrupted by the change in sign of the effective pitch rate of the blade kinematics.
When the blade's effective angle of attack decreases, the leading-edge vortex is forced to pinch-off before it has grown to its full size and strength.
These conditions are known as light stall conditions.
Vertical-axis wind turbines operating with $\lambda>4.5$ are shown in grey as the blade does not exceed its static stall, measured to be $\kindex{\alpha}{ss}=\ang{13.1}$.
The boundary between light and deep dynamic stall is not very sensitive to the exact choice of the static stall angle, as the timescales vary only slightly for a small change in angle at high reduced frequencies.
Note that this map does not account for blade-wake interactions arising from the presence of other blades.
The map is also designed to diagnose the upwind half of the turbine rotation ($0 \leq t/T < 0.5$).
During the downwind half ($0.5 \leq t/T < 1$), the blade may interact with previously formed wakes (\cref{fig:Ctot1}\subf{e-h}).
These wake interaction can trigger the onset of flow separation even if the critical stall angle is not exceeded.
These effects are not included in \cref{fig:DSmap}.

At low tip-speed ratios ($\lambda < 2.5$ for $c/D = 0.2$), vertical-axis wind turbines could achieve the highest power production and might seem attractive at first.
The blade experiences a significant peak in thrust (positive $\kindex{C}{$\theta$}$) in the first quarter of the rotation ($0 \leq t/T <0.25$), which is followed by large excursions in drag (negative $\kindex{C}{$\theta$}$) in the post-stall regime ($0.3 \leq t/T <0.8$).
This range of operating conditions is characterised by the formation of a large and coherent dynamic stall vortex, which interacts with the blade over an extended period of time when it convects downstream along the blade's path at the overlap between the upwind and downwind part of the rotation.
Massive flow separation and blade-vortex interaction lead to important load variations that might challenge the structural integrity of the wind turbine and decreases the appeal of low tip-speed ratio regime.

At intermediate tip-speed ratios ($2.5 \leq \lambda < 4.5$ for $c/D = 0.2$), the main torque production ($\kindex{C}{$\theta$} > 0$) occurs partly during upwind ($0.15 \leq t/T <0.5$), but also partly during downwind ($0.6 \leq t/T < 1$), as the turbine blade has time to recover from the light dynamic stall event that occurs during the upwind part of the motion.
The leading-edge vortex is smaller in size and strength than at low tip-speed ratios and is shed from the blade before it has the change to grow to its full capacity, as the blade starts to pitching down less than three convective times after the blade has exceeded its static stall angle.
The load transients related to flow separation for intermediate tip-speed ratios are \SI{75}{\percent} lower than at low tip-speed ratios.

At high tip-speed ratios ($4.5 \leq \lambda$ for $c/D = 0.2$), the production of torque is significantly reduces and there is no vortex formation on the blade ($\alpha < \kindex{\alpha}{ss}$).

Even though the wind turbine would produce less torque at the intermediate tip-speed ratios than for the low tip-speed ratios, the lighter stall and lower stall-related load fluctuations make this regime the sweet spot for typical vertical-axis wind turbine operation and the preferred solution to the dynamic stall dilemma.
These findings agree with previous studies in literature that found optimal operating conditions to be around a tip-speed ratio $\lambda= 3$ for wind turbines operating under comparable conditions \cite{Rezaeiha2018a,Scheurich2012,Jamieson2011}.

%				--------	  			~			conclusion 	 		~					--------				%

\section{Conclusion}

%				--------	  			~			conclusion 	 		~					--------				%

Time-resolved force and velocity field measurements were conducted on a scaled-down model of a Darrieus H-type wind turbine operating at a chord-based Reynolds number $Re = \num{50000}$, chord-to-diameter ratio $c/D = 0.2$ and tip-speed ratio $\lambda \in [1.2 - 6]$.
The development of dynamic stall on a vertical-axis wind turbine was characterised over this comprehensive operational envelope.

Vertical-axis wind turbines operating conditions are characterised by the tip-speed ratio $\lambda$ and chord-to-diameter ratio $c/D$.
Based on the flow and force data, we categorised the response of the wind turbine blade as a function of the operating conditions into three regimes: no stall, light dynamic stall, and deep dynamic stall.
Low tip-speed ratios ($\lambda < 2.5$ at $c/D = 0.2$) achieve a significant peak in torque production around $\kindex{C}{$\theta$,max} = 1.1$ when a dynamic stall vortex grows during the upwind half of the rotation.
The maximum tangential force coefficient is directly related to the strength of the dynamic stall vortex and both increase with decreasing tip-speed ratio.
When the stall vortex pinches-off before the maximum effective angle of attack is reached at low tip-speed ratios, the blade experiences deep stall which is associated with significant fluctuation of the loads and an asymmetry in the torque production between the upwind and downwind half of the turbine revolution.
At intermediate tip-speed ratios ($2.5 \leq \lambda \leq 4.5$), the maximum angle of attack is reached before the dynamic stall vortex has had enough time to grow to its full potential and the blade experiences light dynamic stall.
The formation of the leading-edge vortex is interrupted by the change in pitch rate, yielding a smaller vortex with a \SI{60}{\percent} reduction in vortex strength and tangential force maximum compared to the low tip-speed ratio cases.
The post-stall load fluctuations are significantly reduced at intermediate tip-speed ratios, with a \SI{75}{\percent} reduction in load transients compared to low tip-speed ratios.
The turbine blade recovers faster from light stall conditions and achieves a second torque production region in the downwind half of its rotation.
Intermediate tip-speed ratios offer a desirable compromise in the dynamic stall dilemma, as they strike a balance between torque production and structural resilience.
At high tip-speed ratios ($\lambda > 4.5$) the blade does not exceed its critical stall angle and dynamic stall does not occur.
Future efforts will focus on developing smart active control strategies that would allow us to mitigate stall related load transients and to tap into the high torque production potential at low tip-speed ratios.

%				--------	  			~		acknowledgements 	 	~					--------				%

\section*{Acknowledgements}

%				--------	  			~		acknowledgements 	 	~					--------				%

This work was supported by the Swiss national science foundation under grant number PYAPP2\_173652.

%				--------	  				~		Appendix 	 	~						--------				%

\appendix
\crefname{appendix}{}{}

%				--------	  				~		Appendix 	 	~						--------				%

\section{Load cell design and calibration}\label{sec:ap_lc}

\begin{figure}[tb!]
\centering
\includegraphics{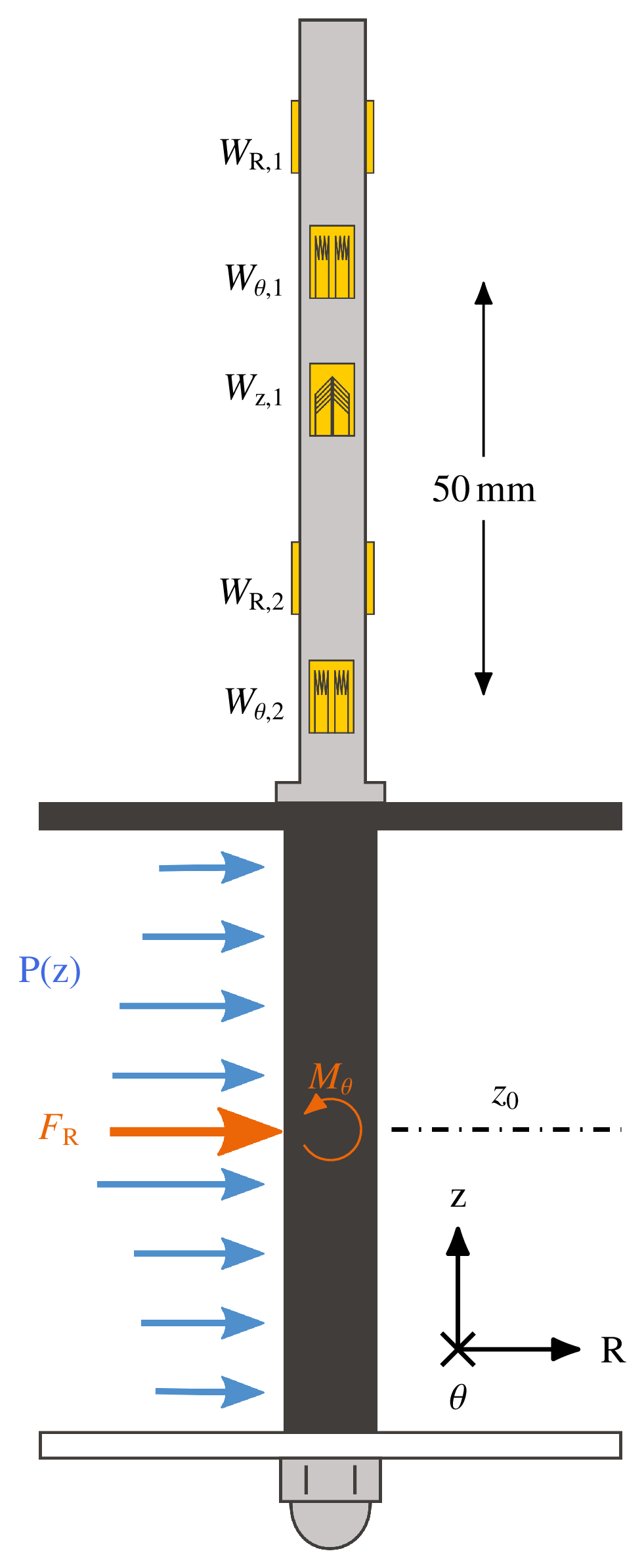}
\caption{Load cell design diagram. Each Wheatstone bridge compose of a pair of strain gauges is indicated by a $\kindex{W}{x}$ where $x$ represents the direction of the measured force or moment axis in the case of $\kindex{W}{z}$. The force distribution exerted by the flow on the wind turbine blade is represented as $P(z)$. In this orientation, the load distribution results in a shear force \kindex{F}{R} and a bending moment \kindex{M}{$\theta$}. The diagram is not to scale.}
\label{fig:lcdia}
\end{figure}

An in-house load cell was built into the wind turbine's blade shaft.
This was achieved with eight double bending strain gauges and two double torsion strain gauges.
Each pair of strain gauges form a full Wheatstone bridge.
Full Wheatstone bridges are twice as sensitive as half-bridges and they compensate for temperature variations originating, e.g. from the electronics hearting up.
A schematic front view of the load cell shows the positions of the Wheatstone bridges $W$ in \cref{fig:lcdia}.
The bridges \kindex{W}{R} and \kindex{W}{$\theta$} respond to forces in the $R$ and $\theta$ directions, and \kindex{W}{z} to moments about the $z$ direction.
An example load distribution $P(z)$ of the flow acting on the turbine blade is shown.
In this example, we want to divide the loading $P(z)$ into the resulting shear force \kindex{F}{R} and the moment \kindex{M}{$\theta$} about the centre of the
blade, to be able to isolate the relevant aerodynamic forces.
This decoupling is achieved by installing the bridge pairs (\kindex{W}{R,1}, \kindex{W}{R,2}) and (\kindex{W}{$\theta$,1}, \kindex{W}{$\theta$,2}) \SI{50}{\milli\meter} apart along the shaft, as indicated in the figure.
%The shear forces \kindex{F}{$\theta$} and \kindex{F}{R} were decoupled from the moments \kindex{M}{R} and \kindex{M}{$\theta$} by installing Wheatstone bridge pairs \SI{50}{\milli\meter} apart along the shaft.
%The Wheatstone bridge pairs are composed of the bending strain gauges and are depicted in \cref{fig:lcdia} as \kindex{W}{R} or \kindex{W}{$\theta$} depending on the direction of the force they respond to.
With this configuration, the bending moment $BM$ in each direction is measured at two positions along the shaft, allowing to solve the equation $BM = F \kindex{l}{z} + M$ for the shear
force $F$ and moment $M$, where $\kindex{l}{z}$ is the distance from the blade's midspan location \kindex{z}{0}.
% This condition is required for the linear regression to be successful at finding the coefficients of the calibration matrix.

The load cell measures the voltages from the five Wheatstone bridges, which are converted into actual forces and moments with the
calibration matrix $R$, following
%lcdia_an
\begin{equation*}
    F= R \cdot V \text{~, or }
\end{equation*}
\begin{equation}
	\begin{bmatrix}
	\kindex{F}{$\theta$} \\
	\kindex{F}{R} \\
	\kindex{M}{$\theta$} \\
	\kindex{M}{R} \\
	\kindex{M}{z}
	\end{bmatrix}
	=
	\begin{bmatrix}
	\kindex{R}{11} & \kindex{R}{12} & \cdots & \kindex{R}{15} \\
	\kindex{R}{21} & \kindex{R}{22} & \cdots & \kindex{R}{25} \\
	\vdots & \vdots & \ddots & \vdots \\
	\kindex{R}{51} & \kindex{R}{52} & \cdots & \kindex{R}{55}
	\end{bmatrix}
	\cdot
	\begin{bmatrix}
	\kindex{V}{1} \\
	\kindex{V}{2} \\
	\kindex{V}{3} \\
	\kindex{V}{4} \\
	\kindex{V}{5}
	\end{bmatrix}
\end{equation}
where $\kindex{F}{$\theta$/R}$ are the forces in the blade's tangential and radial directions, $\kindex{M}{$\theta$/R/z}$ are the moments acting at the blade's half span, quarter-chord position about the tangential, radial, and spanwise axes, $\kindex{V}{i}$ are
the voltages obtained from the $i^{th}$ Wheatstone bridge and $\kindex{R}{ij}$ are the
calibration matrix coefficients obtained from the linear regression.
Note that for the physical interpretation of the current work, \kindex{M}{$\theta$} and \kindex{M}{R} are not needed.
To obtain the calibration matrix $R$, we apply known forces and moments $F^\text{calib}$ and measure the corresponding output voltages $V^\text{calib}$.
It is then possible to determine the inverse of the calibration matrix with
\begin{equation}
    V^\text{calib}=R^{-1} \cdot F^\text{calib} .
    \label{eq:calib_inv}
\end{equation}
The matrix $R^{-1}$ is computed by solving \cref{eq:calib_inv} with a least-squares fit, and the calibration matrix $R$ is simply obtained by inverting $R^{-1}$.

\begin{figure}[tb!]
\centering
\includegraphics{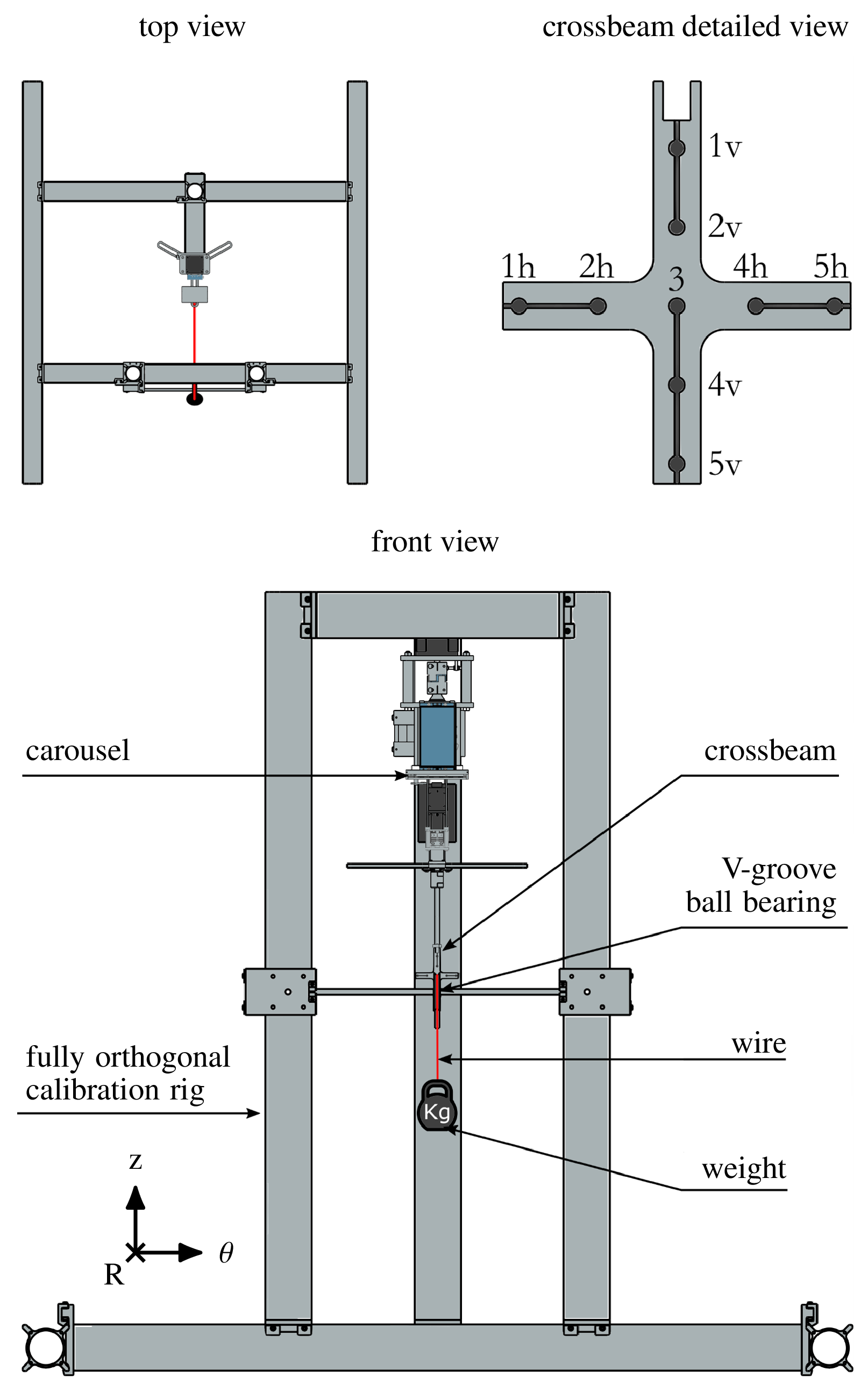}
\caption{Vertical-axis wind turbine load cell orthogonal calibration rig.}
\label{fig:calirig}
\end{figure}

To perform the calibration procedure, the wind turbine is mounted on a
fully-orthogonal test rig.
Loads are applied on the load cell using a v-groove ball bearing or a frictionless pulley, a wire, and a crossbeam mounted on the blade shaft instead of the turbine blade.
The calibration rig is shown in \cref{fig:calirig}.
Loads and moments acting in the desired direction are generated by changing the cross beam orientation relative to the bearing and the force application point in the cross beam.
We apply weights statically using the wire and ball bearing and acquire
measurements for \SI{10}{\second}, to get the mean voltage and standard deviation.
For each position on the crossbeam, we apply nine loading conditions consecutively.
The nine loading conditions are multiples of a unit weight $W \approx \SI{300}{gram}$ that was selected to properly discretise the expected range of forces to be measured by the wind turbine model.
The loading routine was as follows: $0W$, $1W$, $2W$, $3W$, $4W$, $3W$, $2W$, $1W$, $0W$.
This routine also enables us to determine the error related to loading hysteresis.
In total, \num{1350} static measurements are performed to generate the calibration data used in \cref{eq:calib_inv}.

The calibration procedure demonstrated a very satisfactory linearity between measured voltage and applied force, and very low hysteresis ($<\SI{1}{\percent}$).

% -----------------------------------------------
\section{Load cell uncertainty\label{sec:app_uncertainty}}

The load cell uncertainty $\kindex{U}{lc}$ is important to estimate the level of trust we can attribute to the measured forces.
This uncertainty combines both the load cell's precision and accuracy.
The precision is given by the disagreement between repeated measures of the same reference loading condition and is quantified by the standard deviation $\sigma$ of these measurements.
The accuracy defines the error $e$ between the experimental values and the true reference.
The experimental value is determined as the average of the repetitions for a single loading condition.
The combination of the error and the standard deviation yields the uncertainty
\begin{equation}
  \kindex{U}{lc}=\sqrt{\sigma^2+e^2}\quad.
  \label{eq:uncertainty}
\end{equation}

\begin{figure}[t!]
   \centering
   \includegraphics[width=0.5\textwidth]{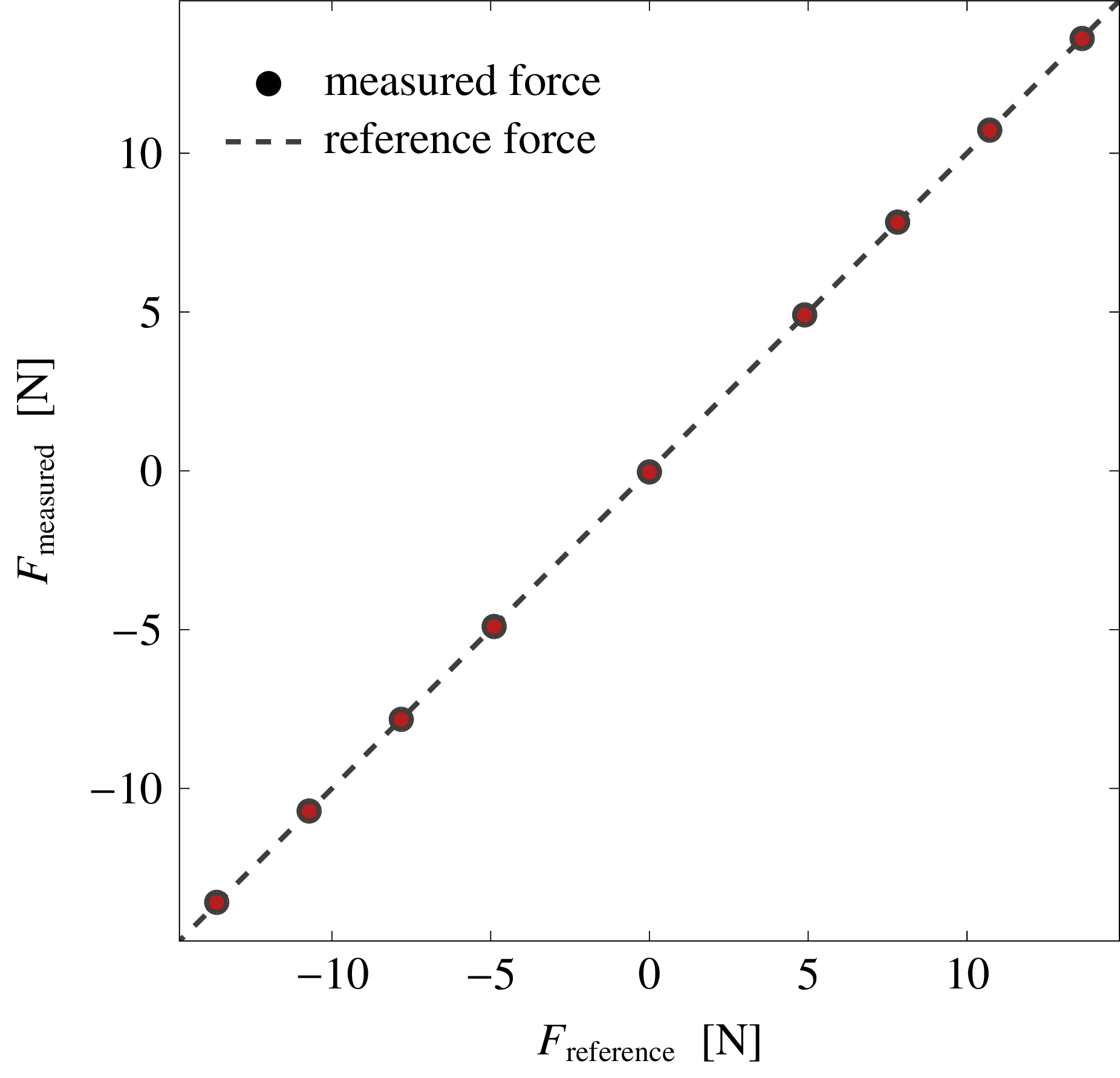}
   \caption{Force measured by the load cell in the tangential $\theta$ direction compared to reference force applied with a static weight. Error bars are very small and hidden by the marker.}
   \label{fig:uncertainty}
\end{figure}

As an example, the calibration results for the tangential force \kindex{F}{$\theta$} are presented in \cref{fig:uncertainty}.
The forces obtained from the voltage measurements multiplied by the calibration matrix \kindex{F}{measured} are shown against the reference forces statically applied with the weights \kindex{F}{reference}.
For each loading condition, the measurement is repeated between \num{50} and \num{100} times.
The diagonal curve indicate the ideal case, where the reference and measured values coincide, and the deviation from this line indicate the error $e$.
The uncertainty is computed for each loading conditions with
\cref{eq:uncertainty} and averaged.

Following this procedure, the uncertainty for all loads and moments was determined, and are listed in \cref{tab:uncertainty}.
The values reveal an overall low experimental uncertainty, below \SI{5}{\percent} of the dynamic range, and confirm the reliability of the load cell results.
\begin{table}[t!]
  \centering
  \begin{tabular}{c|c|c}
    \textbf{Component} & \textbf{Max range} & \textbf{Uncertainty} \\
    \hline
    \kindex{F}{$\theta$} & $\pm$\SI{50}{\newton} & $\pm$ \SI{0.06}{N} \\
    \kindex{F}{R} & $\pm$\SI{50}{\newton} & $\pm$ \SI{0.20}{\newton}\\
    \kindex{M}{x} & $\pm$\SI{0.5}{\newton\meter} & $\pm$ \SI{0.009}{\newton\meter}\\
    \kindex{M}{y} & $\pm$\SI{0.5}{\newton\meter} & $\pm$ \SI{0.006}{\newton\meter}\\
    \kindex{M}{z} & $\pm$\SI{0.5}{\newton\meter} & $\pm$ \SI{0.005}{\newton\meter}\\
  \end{tabular}
  \caption{Load cell uncertainty for each load and moment.}
  \label{tab:uncertainty}
\end{table}
% -----------------------------------------------

\section{Force offset}\label{sec:ap_offset}

\begin{figure}[]
	\centering
	\includegraphics{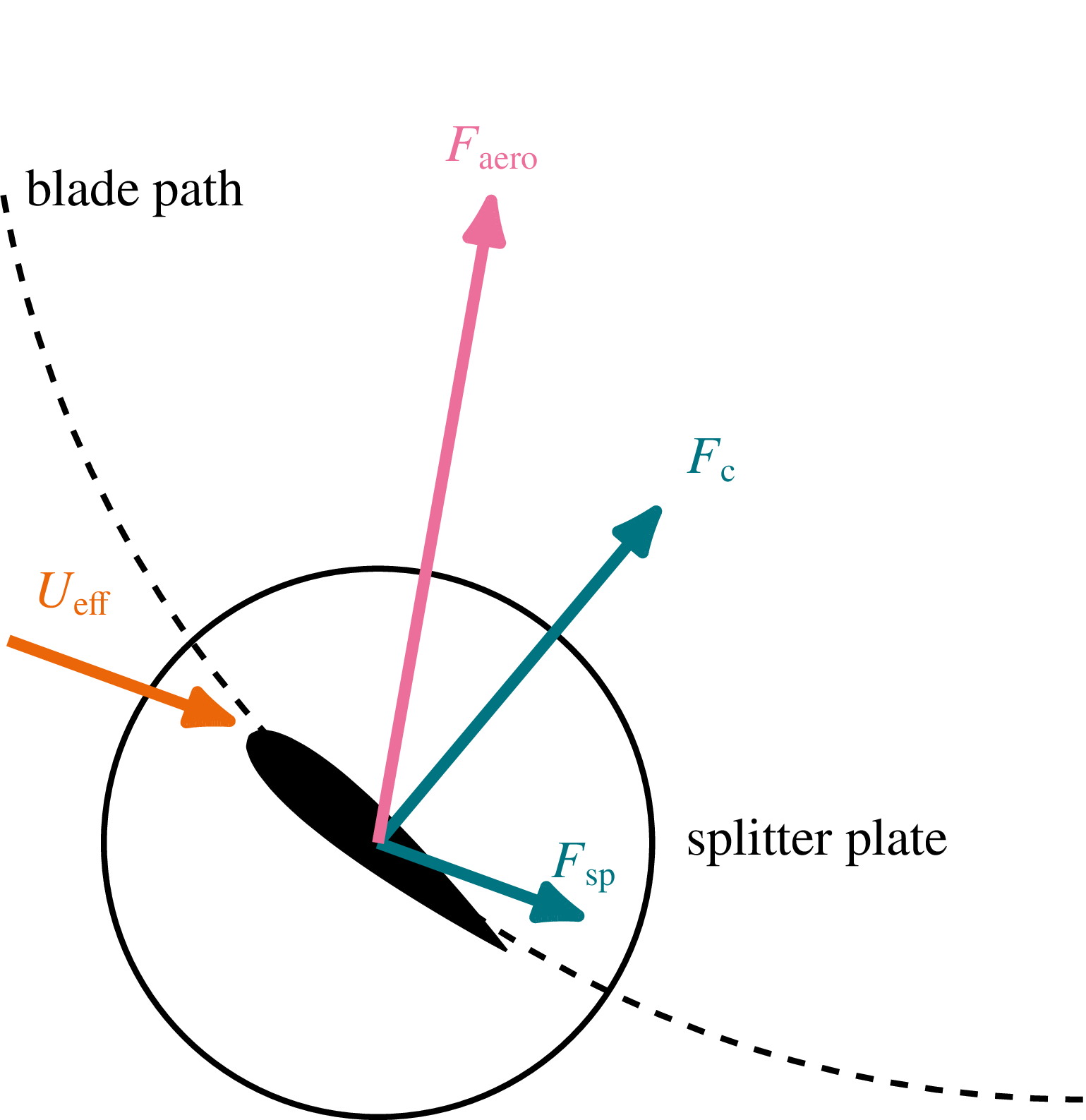}
	\caption{Force offset diagram.}
	\label{fig:fdiagram}
\end{figure}

We wish to investigate the aerodynamic forces \kindex{F}{aero} experienced by the wind turbine blade.
The built-in load cell captures two additional force contributions during wind turbine experimentation in the water channel: the centripetal force $\kindex{F}{c}$ related to the mass below the load cell following a circular path and the aerodynamic splitter plate force $\kindex{F}{sp}$ (\cref{fig:fdiagram}).
These two force components were isolated, measured, and subtracted from the raw force data to obtain the results presented in this paper.

\begin{figure}[]
	\centering
	\includegraphics[width=\linewidth]{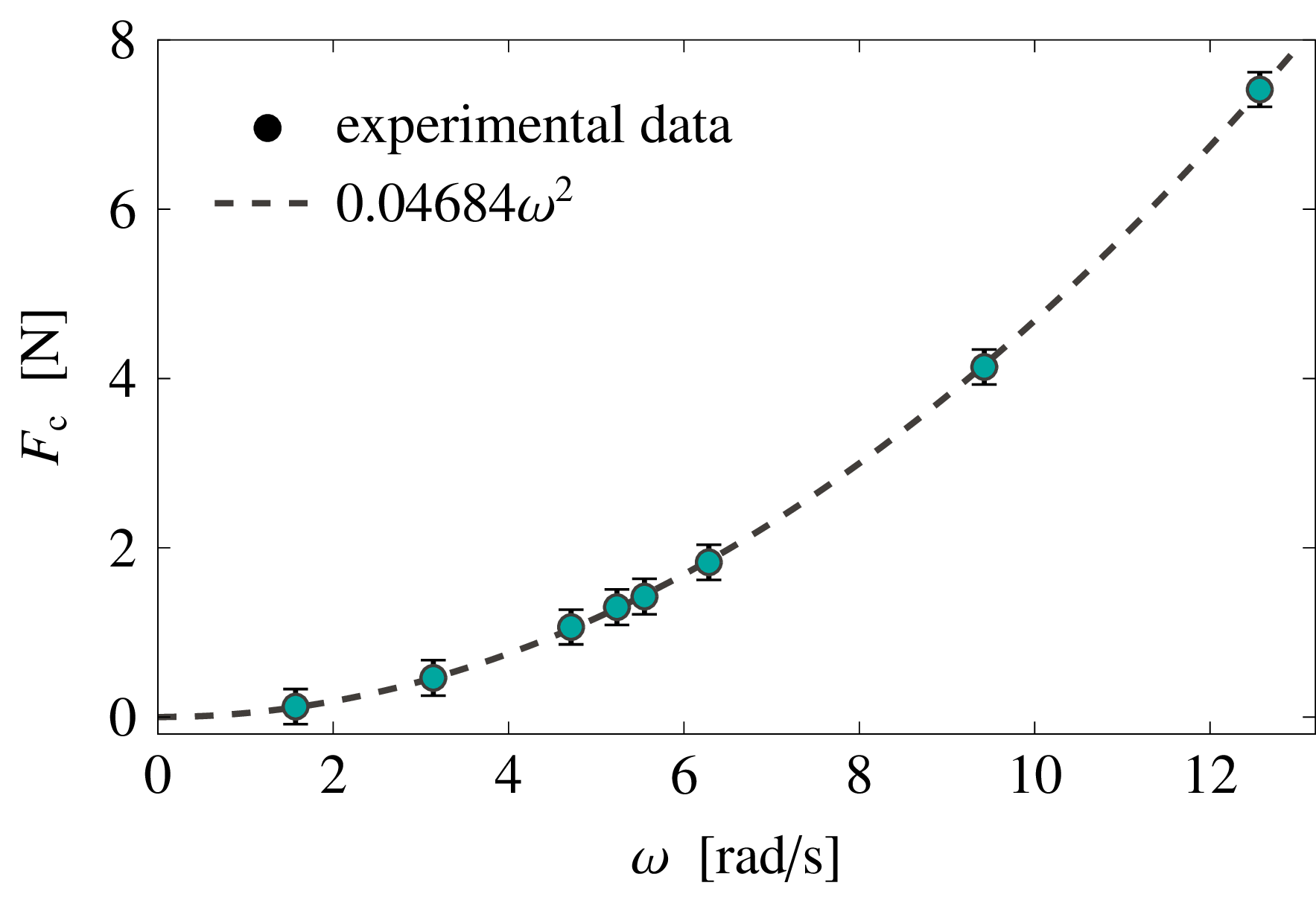}
	\caption{Centripetal force \kindex{F}{c} measurement and modelling. The scattered points represent the averaged experimental measurements.
  Error bars represent the absolute error computed in \cref{sec:app_uncertainty}.
  The dashed line is the fitted polynomial function used to model the centripetal force.}
	\label{fig:ApFc}
\end{figure}

The centripetal force was measured by rotating the wind turbine blade in air at eight different rotational frequencies spanning the range of expected experimental rotational frequencies.
For each rotational frequency, the centripetal force was measured for
\SI{30}{\second}.
The mean centripetal force is presented as a function of the rotational frequency in \cref{fig:ApFc}.
The error bars represent the absolute error computed in \cref{sec:app_uncertainty}.
The centripetal force evolves with $\kindex{F}{c} = mR\omega^2$, where $m$ is the mass below the load cell, $R$ the radius of the blade's circular path, and $\omega$ is the turbine's rotational frequency.
A second-order polynomial fit was used to determine $mR$.
The experimental data fit yielded an expression for the centripetal force $\kindex{F}{c} = 0.047 \omega^2$, as shown in \cref{fig:ApFc}.

\begin{figure}[]
	\centering
	\includegraphics[width=\linewidth]{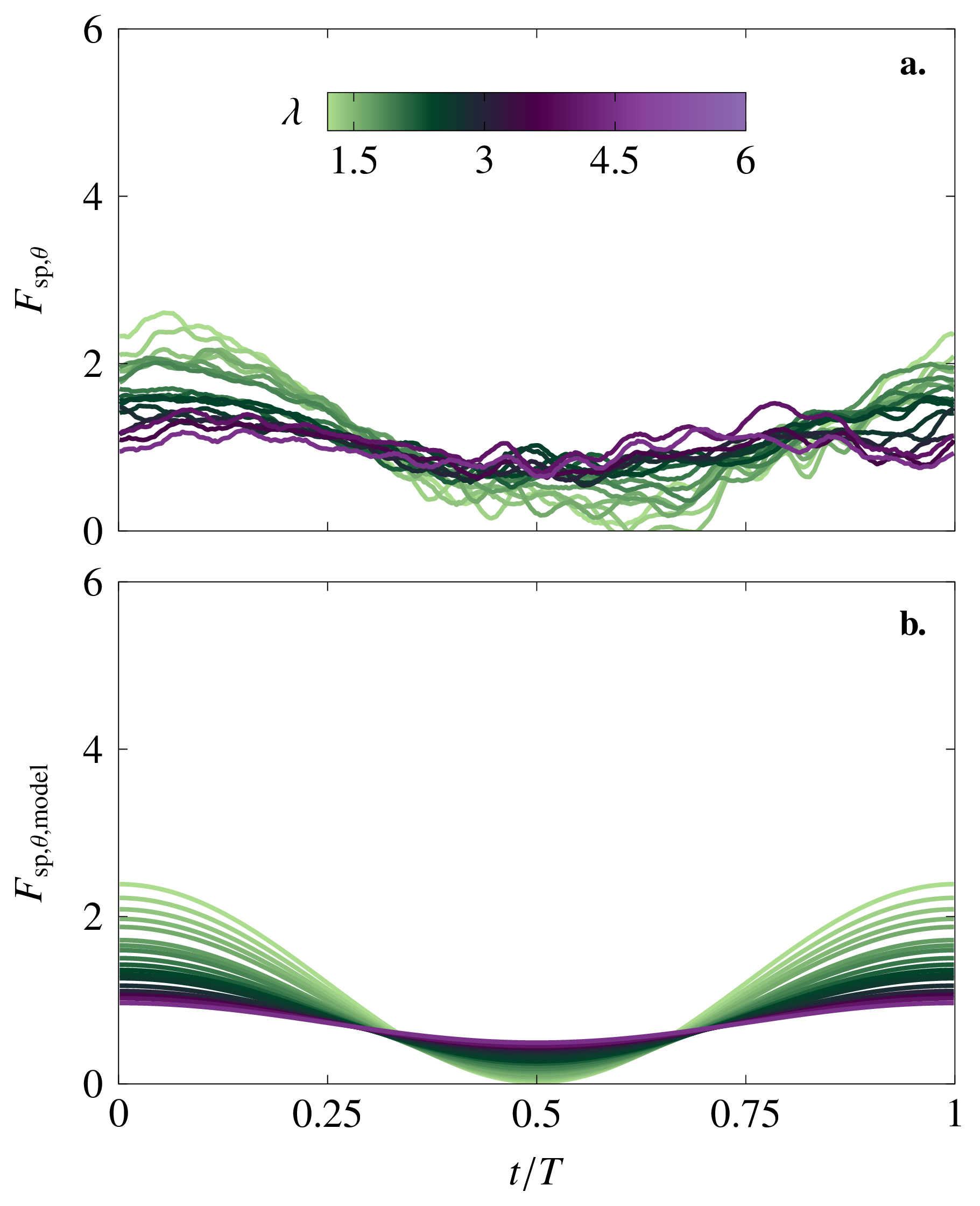}
	\caption{Tangential force related to the splitter plates (a) experimental measurement and (b) model.}
	\label{fig:ApFt}
\end{figure}

The aerodynamic forces related to the presence of the splitter plates \kindex{F}{sp} was measured by operating the vertical-axis wind turbine in the water channel without the NACA0018 blade.
Instead, a $\kindex{D}{cyl} = \SI{12}{\milli\meter}$ diameter and $\kindex{L}{cyl} = \SI{150}{\milli\meter}$ long cylinder was placed in between the two splitter plates to hold them in place.
The aerodynamic forces generated by the cylinder \kindex{F}{cyl} were offset using $\kindex{F}{cyl} = 0.5\kindex{C}{D,cyl}\rho \kindex{U}{eff}^2 \kindex{D}{cyl} \kindex{L}{cyl}$, where \kindex{C}{D,cyl} was obtained from \cite{Roshko1961}, and \kindex{U}{eff} is calculated using \cref{eq:Ueff}.
The inertial forces from the cylinder setup were also measured and subtracted from the splitter plate measurements.
The splitter plate force \kindex{F}{sp,R} in the radial direction was small, below $\SI{5}{\percent}$ of the expected dynamic load during vertical-axis wind turbine experimentation.
This value is considered to be within the error margin of the aerodynamic measurements and is ignored.
The temporal evolution of \kindex{F}{sp,$\theta$} in the tangential direction for a wide range of tip-speed ratios is presented in \cref{fig:ApFt}a.
The range of the vertical axis is shown to represent the typical range of forces acquired in the tangential direction during wind turbine experiments.
Note that the maximum force in this direction reach values above \SI{10}{\newton}.
As expected, \kindex{F}{sp,$\theta$} is maximum at $t/T = 0$ when the splitter plates are facing the flow and the effective velocity $\kindex{U}{eff}$ is maximum.

The splitter plate forces are modelled using a typical drag coefficient formulation:
\begin{equation}
	\kindex{F}{sp,$\theta$} = \kindex{C}{D,sp}\rho \kindex{U}{eff}^2 \pi \kindex{R}{cyl}^2
\end{equation}
where \kindex{R}{cyl} is the radius of the splitter plates and \kindex{C}{D,sp} is the drag coefficient to be determine by fitting the experimental data.
A polynomial fit yielded a drag coefficient of $\kindex{C}{D,sp} = 0.09$.
The modelled splitter plate forces in the tangential direction are presented for the same range of tip-speed ratio in \cref{fig:ApFt}b.
The model represents the experimental data best in the first and last quarter of the turbine rotation when the force is largest.
The force minimum is slightly delayed and extended for the experimental data compared to the modelled forces around $t/T=0.5$.
The delay is arguably related to wake delay and vortical structures interacting with the cylinder and splitter plates in the region.
The small difference between the model and experimental results is deemed acceptable because the splitter plate influence is below \SI{1}{\newton} when $0.25 \leq t/T < 0.75$.
The model offers a satisfactory representation of the splitter plate interaction.

%				--------	  			~			Bibliography 	 	~					--------				%

\bibliography{extracted.bib}

%				--------	  			~			Bibliography 	 	~					--------
\end{document}